\documentstyle[12pt, epsf]{article}

\begin{document}
\renewcommand{\theequation}{\thesection.\arabic{equation}}
\thispagestyle{empty} \vspace*{-1.5cm} {\small hep-th/0002044}
\hfill {\small SU-ITP 99-14}
%
\hfill {\small KUL-TF-2000/03}
\\[8mm]

\setlength{\topmargin}{-1.5cm} \setlength{\textheight}{22cm}
\begin{center}
{\large TASI lectures on the Holographic Principle }
\\

\vspace{1 cm} {\large  Daniela Bigatti }
\\ Institute of Theoretical Physics,
KU Leuven,
\\ B-3001 Heverlee, Belgium 
\\ \vspace{0cm}

\vspace{1 cm} {\large  Leonard Susskind }
\\ Department of Physics,
Stanford University,
\\ Stanford CA 94305-4060
\\ \vspace{0cm}

\begin{abstract}
These TASI lectures review the Holographic principle. The first
lecture describes the puzzle of black hole information loss that
led to the idea of Black Hole Complementarity and subsequently to
the Holographic  Principle itself. The second lecture discusses
the holographic entropy bound in general space-times. The final
two lectures are devoted to the ADS/CFT duality as a special case
of the principle. The presentation is self contained and
emphasizes the physical principles.   Very little technical
knowledge of string theory or supergravity is assumed.
\end{abstract}

\vspace{1cm} {\it August 1999}
\end{center}

\setcounter{equation}{0}
\section{Black Hole Complementarity}

New scientific ideas are usually characterized by simple
organizing principles that can be expressed in a phrase or two.
The invariance of the speed of light,  the equivalence principle
the uncertainty principle  and survival of the fittest are famous
examples. Is there a comparable simple summary of the new
principles which our science is now uncovering? Some people think
it is supersymmetry, others think it is duality. But the real
world is not supersymmetric, nor is it known to have dual
descriptions in any deep sense. My own view is that the lasting
idea will be the {\it{holographic principle}} \cite{1}\cite{2},
the assertion that the number of possible states of a region of
space is the same as that of a system of binary degrees of freedom
distributed on the boundary of the region. The number of such
degrees of freedom is not indefinitely large but is bounded by the
area of the region in Planck units. These lectures are about the
motivations and evidence for this principle.

The holographic principle grew out of the deep insights of
Bekenstein \cite{3} and Hawking \cite{4} in the  70's. In
particular Hawking raised a very profound question concerning the
consistency of gravitation and the usual operational principles of
quantum mechanics \cite{5}. To state the paradox clearly it is
useful to think of a black hole as an intermediate state in a
scattering process. Particles, perhaps in the form of stars,
galaxies or just ordinary quanta come together in an initial state
$|in \rangle$. A black hole forms and evaporates leaving outgoing
photons, gravitons neutrinos and other quanta. No energy is lost
in the process so there are no unaccounted for degrees of freedom
in the final state. According to the usual rules, such a process
is described by a unitary scattering matrix $S$.
\begin{equation}
|out\rangle =S |in\rangle
\end{equation}
Since $S$ is unitary we can also write
\begin{equation}
|in\rangle =S^{\dagger} |out\rangle
\end{equation}
In other words it must be possible to recover the initial quantum
state from the final state in a unique way. However, Hawking gave
arguments, that appeared to many as completely persuasive, that
information is irretrievably lost when matter falls behind the
horizon of the black hole. Thus, from an operational point of
view, the rules of quantum mechanics as set out by Dirac would
have to be modified as collision energies approach and exceed the
Planck energy. In particular  the conventional $S$ matrix would
not exist.

Not everyone believed Hawking's arguments \cite{6} \cite{7}.
{\it{Black hole complementarity}} \cite{8} and the holographic
principle \cite{1} \cite{2} are counter-proposals that preserve
intact the general principles of quantum mechanics but question
some of the naive beliefs about locality and the objectivity and
invariance of space-time events.
\bigskip

{\it{The Schwarzschild Black Hole}}

To understand the issues we will need to review the geometry of
black holes. There are many kinds of black holes in string theory
but we will confine our attention to the usual $3+1$ dimensional
Schwarzschild case.

The ordinary Schwarzschild black hole is described by the metric
\begin{equation}
ds^2=\left( 1-{2MG \over r}\right)dt^2 -\left( 1-{2MG \over
r}\right)^{-1} dr^2 - r^2 d \Omega^2
\end{equation}
$M,G$ and $d\Omega^2$ are the black hole mass, the gravitational
constant and the metric of a unit 2-sphere. The curvature
singularity at $r=0$ will not concern us but the coordinate
singularity at the Schwarzschild radius $r = 2MG$ defines the all
important horizon. Despite its singular importance, the horizon is
not a mathematical singularity of the geometry, at least in the
usual sense. To see that let us concentrate on the "near  horizon
limit". We consider a small angular region near a point on the
horizon. Define
\begin{equation}
y=r-2MG
\end{equation}
For $y<< 2MG$ the metric has the form
\begin{equation}
ds^2 = {y \over 2MG}dt^2 -{2MG \over y}dy^2 - dx^i dx^i
\end{equation}
where $dx^i$ is an element of length in the two dimensional plane
tangent to the horizon. Now define
\begin{eqnarray}
\rho &=& \sqrt{8MGy} \cr \omega&=& {t \over {4MG}}
\end{eqnarray}
and the metric takes the form
\begin{equation}
ds^2=\rho^2 d\omega^2 -d\rho^2 -dx^idx^i
\end{equation}
Expression (1.7) is the metric of ordinary Minkowski space in
hyperbolic polar coordinates. If we define
\begin{eqnarray}
X^+ &=& \rho e^{\omega} \cr X^- &=& -\rho e^{-\omega}
\end{eqnarray}
the metric  becomes
\begin{equation}
ds^2=dX^+ dX^- -dx^i dx^i
\end{equation}
which is the standard light cone form of the Minkowski metric.
>From this fact it is apparent that the horizon is not singular.

The relation between the flat minkowski coordinates $X^\pm$
and the Schwarzschild coordinates $r,t$ is shown in figure(1) for
the region outside the horizon. The entire horizon $r=2MG$ is
mapped to the point (2D-surface) $X^+=X^- = 0$. The extended
horizon is defined by the 3 dimensional surface $X^- = 0$. Notice
that a signal originating from a point behind the horizon, $X^- >
0$ can never escape to the outside, $X^- < 0$. For the region
$X^+>0$, the extended horizon coincides with the asymptotic
limiting value of Schwarzschild time $t= \infty$. Although the
flat Minkowski coordinates only describe the near horizon region,
a generalization to Kruskal-Szekeres (KS) coordinates covers the
whole black hole space-time. Suppressing the angular coordinates
$\Omega$ the KS metric has the form
\begin{equation}
ds^2=F(X^+X^-)dX^+ dX^-
\end{equation}
where $F \to 1$ for $X^+X^- \to 0$ and
\begin{equation}
F \to {16M^2 G^2 \over \rho^2}
\end{equation}
for $X^+X^- \to \infty$. Equation (1.11) insures that the metric
far from the black hole tends to flat space
\begin{equation}
ds^2 \to dt^2-dr^2 -r^2 d\Omega^2
\end{equation}

In KS coordinates the singularity at $r=0$ is defined by the
{\it{space-like}} surface
\begin{equation}
X^+X^- =M^2G^2
\end{equation}
In figure (2) the geometry of the black hole is shown for the
region $X^+>0$.

Now consider a particle trajectory which begins outside the black
hole, falls through the horizon and eventually hits the
singularity as shown in figure (3). In Schwarzschild coordinates
the particle does not cross the horizon until infinite time has
elapsed. Thus from the viewpoint of an observer outside the black
hole, the particle asymptotically approaches the horizon, but
never crosses it. Indeed, all the matter which collapsed to form
the black hole never crosses the horizon in finite Schwarzschild
time. Classically it forms progressively thinner layers which
asymptotically approach the horizon.

On the other hand, from the point of view of a freely falling
observer accompanying the infalling particle the horizon is
crossed after a finite time. In fact from figure {3} it is obvious
that nothing special happens to the infalling matter at the
horizon. This discrepancy is the first instance of an
under-appreciated complementarity or relativity between the
descriptions of matter by external and infalling observers.
\bigskip

{\it{Penrose Diagrams}}

Penrose diagrams provide an intuitively clear way to visualize the
global geometry of black holes. They are especially useful for
spherically symmetric geometries.  The Penrose  diagram describes
the $r,t$ plane. Here are the rules for a Penrose diagram.

1.  Use coordinates which map the entire geometry to a finite
portion of the plane.

2. The coordinates should be chosen so that radial light rays
correspond to line oriented at $\pm 45$ degrees to the vertical.

As an example the Penrose diagram for flat space is shown in
figure (4). The vertical axis is the spatial origin at $r=0$ and
the point labeled $r = \infty$ represents the asymptotic endpoints
of space-like lines. The points $t = \pm \infty$ are the points
where time-like trajectories begin and end. Light rays enter from
past null infinity, $\Im^-$  and exit at future null infinity,
$\Im^+$.

The Penrose diagram for the Schwarzschild geometry is shown in
figure (5).  As we will see the regions III and IV are unphysical.
Region I is the outside of the black hole and like flat Minkowski
space it has space-like, time-like and null infinities. Obviously
future directed time-like or light-like trajectory that begins in
region II will collide with the singularity. Thus region II is
identified as being behind the horizon. The extended horizon (from
now on called the horizon) is the light-like line $t= \infty$.

A real black hole must be formed in a collapse. Thus in the remote
past there is no black hole and the geometry should resemble the
lower portion of figure (4). At late times the black hole has
formed and the geometry should resemble figure (5). Thus the
Penrose diagram for the collapse looks is shown in figure (6).
\bigskip

{\it{Black Hole Thermodynamics}}

It is well known that black holes are thermodynamic objects
\cite{3} \cite{4} \cite{9}. In addition to their energy, $M$ they
have a temperature and entropy. To understand this we need to
study the behavior of quantum fields in the near horizon geometry.
We will see later that quantum field theory can not really be an
adequate description of a world including gravity but it is a
starting point which will allow us to abstract some important
lessons.

As we have seen, the near horizon geometry is just  Minkowski
space described in hyperbolic polar coordinates. In particular the
portion of the near horizon region $(X^+X^-<0)$ outside the black
hole is called Rindler space.

The usual time coordinate of Minkowski space is $x^0={{X^+ +
X^-}\over 2}$ and conjugate to it is the momentum component $p_0$.
However, $p_0$ is not the energy or Hamiltonian appropriate to the
study of black holes by distant observers. For such observers the
natural time is the Schwarzschild time $t = 4MG \omega$. The
conjugate Hamiltonian which represents the energy or Mass of the
black hole is
\begin{equation}
H_t = {1 \over {4MG}} H_{\omega} ={i \over {4MG}}\partial_{\omega}
\end{equation}
where $H_{\omega}$ is a dimensionless Hamiltonian conjugate to the
dimensionless Rindler time $\omega$.

An observer outside the horizon has no access to the degrees of
freedom behind the horizon. For this reason all observations can
be described in terms of a density matrix $\Re $ obtained by
tracing over the degrees of freedom behind the horizon \cite{9}.
To derive the form of the density matrix for external observations
we begin with the Minkowski space vacuum. The coordinates of
Minkowski space are
\begin{eqnarray}
x^0&=&(X^+ +X^-)/2 \cr x^3&=&(X^+ -X^-)/2
\end{eqnarray}
and the horizon coordinates $x^i$. The instant of Rindler time
$\omega=0$ coincides with the half-surface
\begin{eqnarray}
x^0&=&0 \cr x^3 &>& 0
\end{eqnarray}
The other half of the surface $x^3<0$ is behind the horizon and is
to be traced over.

Let us consider a set of quantum fields labeled $\phi$. To specify
the field configuration at $x^0=0$ we need to give the values of
$\phi$ on both half-surfaces. Let $\phi_I$ and $\phi_F$ represent
the field configurations for $x^3>0$ and $x^3<0$ respectively. A
quantum state is represented by a wave functional
\begin{eqnarray}
\Psi(\phi)= \Psi(\phi_I,\phi_F)
\end{eqnarray}

We use the standard Euclidean  Feynman path integral formula to
compute $\Psi$.
\begin{eqnarray}
\Psi(\phi_I,\phi_F)=\int d\phi \exp{-S}
\end{eqnarray}
where the path integral is over all fields in the future half
space $ix^0>0$ with boundary condition $\phi = (\phi_I, \phi_F)$
at $x^0=0$.

The trick to compute the density matrix $\Re $ is to divide the
upper half plane $ix^0>0$ into infinitesimal angular wedges as in
figure (7). The path integral can then be evaluated in terms of a
generator of angular rotations. This generator is nothing but
$iH_{\omega}$. Thus the expression for the Minkowski vacuum is
\begin{equation}
\Psi(\phi_F,\phi_I) = \langle \phi_F|\exp({-H_{\omega}\pi})|
\phi_I  \rangle
\end{equation}
In other words the Minkowski vacuum wave functional is a
transition amplitude for elapsed Euclidean time $\pi$.

Now consider the density matrix given by
\begin{equation}
\Re = \int d\phi_F \Psi^*(\phi_F,\phi'_I)\Psi(\phi_F,\phi_I)
\end{equation}
Using eq.(1.19)and the completeness of the states $\langle
\phi_F|$ gives
\begin{equation}
\Re = \langle \phi'_I |\exp({-2\pi H_{\omega})} | \phi_I\rangle
\end{equation}
or more concisely
\begin{equation}
\Re = \exp (-H_{\omega}/T_{\omega})
\end{equation}
with $T_{\omega}=1/2\pi$.

Equation (1.22) is has the remarkable property of being a thermal
density matrix for temperature $T_{\omega}$. Notice that the
derivation is exact and in no way relies on the free field
approximation. It is valid for any quantum field theory  for any
strength of  coupling.

The temperature $T_{\omega}=1/2\pi$ does not have the usual
dimensions of energy. This is because the Rindler time and
therefore the Rindler Hamiltonian is dimensionless. To convert to
a proper temperature with dimensions of energy we consider the
proper time interval corresponding to an interval $d \omega$. From
eq.(1.7)
\begin{equation}
d s=\rho d \omega
\end{equation}
Thus an observer at distance $\rho$ from the horizon converts from
dimensionless quantities using the conversion factor $\rho$. The
proper temperature at distance $\rho$ is  given by
\begin{equation}
T(\rho)={1\over \rho} T_{\omega} ={1 \over {2 \pi \rho}}
\end{equation}

In this way we arrive at the important conclusion that an observer
outside a black hole but in the near horizon region will detect a
temperature that varies as the inverse distance from the horizon
\cite{9}.

Next consider the temperature as measured by a distant observer
asymptotically far from the black hole. The proper time variable
for such an observer is the Schwarzschild time $t = 4MG \omega$.
Thus such distant observers measure temperature
\begin{equation}
T_H={T_{\omega} \over {4 \pi MG}}
\end{equation}
This is the Hawking temperature \cite{4} of the black hole. It
represents the true thermodynamic temperature of an isolated black
hole.

The thermodynamic relation between temperature and mass (energy)
allow us to compute an entropy for the black hole. Using
\begin{equation}
dM=T dS
\end{equation}
we find
\begin{equation}
S=4\pi MG
\end{equation}
or in terms of the horizon area $A$
\begin{equation}
S={A \over 4G}
\end{equation}

Equation (1.28) is far more general than the derivation given
here. It applies to every kind of black hole, be it rotating,
charged or in arbitrary dimensions. In the general $(d+1 )$
dimensional case the concept of two dimensional area only needs to
be replaced by the $(d-1)$ dimensional measure of the horizon
which we continue to call area.

\bigskip

{\it{The Thermal Atmosphere}}

Because the region above the horizon has a non-vanishing
temperature, it has a kind of thermal atmosphere \cite{10}
consisting of thermally excited quanta. In regions where the field
theory is weakly coupled the thermal atmosphere consists of
ordinary black body radiation. Some of these quanta have
sufficient energy to escape the gravitational pull of the black
hole and appear as Hawking radiation. However, for a large black
hole, this process is very slow.  The equilibrium approximation
for the thermal atmosphere of the near horizon region  is a very
good one.

The thermal atmosphere contributes to the entropy of the black
hole \cite{11}. Let us consider the ordinary quantum fields of the
standard model or its suitable generalization. For simplicity lets
ignore the interactions as well as masses. The entropy stored in
the shell between $\rho$ and $\rho + d \rho$ for free massless
fields is given by
\begin{equation}
{dS \over {d\rho d^2x^i}}=cT^3
\end{equation}
where $c$ constant proportional to the effective number of
massless fields at that temperature. Using $T=1/{2 \pi \rho}$ we
find
\begin{equation}
S \sim A \int {d \rho \over \rho^3}
\end{equation}

Evidently if this formula made sense all the way to $\rho =0$ the
entropy of the black hole would be infinite. But since we know
that the entropy is $A/4G$ the field theory description must break
down at some small $\rho_0$. In this case the entropy in the
thermal atmosphere of ordinary quanta will be
\begin{equation}
S \sim Ac/{\rho_0^2}
\end{equation}
Since the total black hole entropy is $A/4G$ the contribution from
the thermal atmosphere must be less than this. Accordingly
\cite{11} $\rho_0$ can not be smaller than $\sim G^{1/2}$.

Perhaps a more illuminating way to express this is to say that the
number of effective degrees of freedom must tend to zero as the
Planck temperature is approached \cite{12}. In conventional
quantum field theory  the number of effective degrees of freedom
is a non-decreasing function of temperature. The finiteness of
black hole entropy is  the first evidence that quantum field
theory overestimates the number of independent degrees of freedom.

It is not too surprising that quantum field theory has too many
degrees of freedom at short distances to describe a world with
gravity. The non-renormalizability of quantum gravity has led to
many  suggestions of a Planck scale cutoff over the years. Roughly
speaking, the idea was that there should be about 1 binary degree
of freedom per Planck volume. What we will see in the following is
that this idea still vastly overestimates the number of degrees of
freedom. The correct reduction in the number of degrees of freedom
is that there is no more than $1/R$ degrees of freedom per Planck
volume where $R$ is infrared cutoff radius, that is, the size of
the spatial region being studied.
\bigskip

{\it{The Quantum Xerox Principle}}

The Holographic Principle represents a radical departure from the
principles of local quantum field theory. In order to understand
why we are driven to it we need to follow Hawking's original
arguments about the loss of quantum coherence in black hole
processes. The argument as I will present it is based on a
principle that I call the {\it{quantum Xerox principle}}. It
prohibits the existence  of a machine which can duplicate the
information in a quantum system and in so doing, produce two
copies of the original information. To illustrate an example,
consider a two-state system with states $|u\rangle$ and
$|d\rangle$.  We will call the system  a q-bit. The general state
of the q-bit is
\begin{equation}
|\psi\rangle =  a |u\rangle + b |d\rangle
\end{equation}

Now assume we had a machine which could clone the q-bit and
duplicate a second q-bit in the same state. We can express this by
\begin{equation}
|\psi\rangle  \to |\psi\rangle |\psi\rangle
\end{equation}
For example
\begin{eqnarray}
|u\rangle &\to& |u\rangle |u\rangle \cr |d\rangle  &\to& |d\rangle
|d\rangle
\end{eqnarray}

Suppose a q-bit  in the quantum state $|u\rangle + |d\rangle$ is
fed into the machine. The output of the machine is
\begin{eqnarray}
\left\{ |u\rangle +|d \rangle \right\} \otimes \left\{ |u\rangle +
|d \rangle \right\} = |u  \rangle & \otimes & |u  \rangle \cr
+|d  \rangle & \otimes & |d  \rangle \cr
+ |d \rangle & \otimes & |u  \rangle \cr + |u  \rangle & \otimes &
|d  \rangle
\end{eqnarray}
However this is inconsistent with the most basic principle of
quantum mechanics, the linearity of the evolution of state
vectors. Linearity together with eq. (1.34)   requires
\begin{eqnarray}
|u\rangle +|d\rangle \to |u\rangle \otimes |u\rangle +|d\rangle
\otimes |d\rangle
\end{eqnarray}
In this way we see that the principles of quantum mechanics forbid
the duplication of quantum information. What has all this to do
with black holes?

Consider the following thought experiment \cite{13}. A black hole
is formed as in figure (6). Before the black hole has a chance to
evaporate a q-bit is thrown in. According to the observer who
falls with the q-bit, the information at a later time will be
localized behind the horizon at point (a) in figure (8). On the
other hand an observer outside the horizon eventually sees all of
the energy returned in the form of Hawking radiation. In order
that the usual laws of quantum mechanics are satisfied for the
outside observer, the q-bit of information must be found in the
state of the outgoing evaporation products localized at point (b)
in figure (8).  Since there can not be two copies of the same
information it would seem that either the infalling  observer or
the outside observer must experience a violation of the known laws
of nature. Either the horizon is not such a benign place as we
thought ($\sim$ Minkowski space) and infalling matter is severely
disrupted or else the outside observer experiences a loss of
information in contradiction with quantum principles!

The principle of Black Hole Complementarity flatly denies that
either of these undesirable things happens. According to this
principle no real observer ever detects a violation of the usual
laws of nature. External observations are assumed to be consistent
with a description in which infalling information is absorbed,
thermalized near the hot horizon and returned in the form of
subtle correlations in the Hawking radiation. Furthermore,
infalling observers detect nothing unusual at the almost flat
horizon and only experience violent effects as the singularity is
approached. Reconciliation of these two facts will require that we
radically modify our naive ideas of locality so that the
space-time location of an event loses its invariant significance
and becomes a relative concept.

As we have seen, quantum mechanics forbids information cloning.
Let us take  that to mean that  no real observer is ever allowed
to detect duplicate information. The outside observer has no
problem with this since she can never detect signals from behind
the horizon. However, it is more subtle to argue that observers
behind the horizon can never detect duplicate information. Here is
how it might happen:

An observer, $O$, stationed outside the horizon in figure (9)
collects information stored in the Hawking radiation. After some
time she has collected the information stored in the infalling
q-bit. At that time, she jumps into the black hole, carrying the
information to point (c) behind the horizon . Now there are two
copies of the q-bit behind the horizon, one at (a) and one at (c).
A signal from (a) to (c) can reveal that information has been
duplicated. In fact we will argue that there is a {\it{quantum
Xerox censorship}} mechanism which always prevents this from
happening. To understand it we need one more concept.

\bigskip

{\it{ Information Retention Time}} \\ 
Consider a conventional complex
system such as a piece of coal. Suppose the coal begins in its
ground state and is heated by shining a laser beam on it. As its
temperature rises it begins to glow and emit thermal radiation.
Assume the laser beam is modulated so that it can convey
information and that it sends in a bit.

Let $S$ be the maximum entropy that the coal is heated to before
being allowed to cool back to its ground state. By the time it
does cool, all the information in the laser beam has been returned
in the almost thermal radiation. An interesting question is how
many photons are involved in carrying out the single bit. The
answer has been given in a paper by Don Page \cite{14}. The number
of photons that have to be measured in order to collect a single
bit is of order $S/2$. This is roughly half the photons that will
be emitted. Another way to s say it is that no information can be
retrieved until the coal has cooled to the point where its entropy
is about half its maximum value.

Given the luminosity, this restriction on collecting information
from thermal radiation can be translated to a time scale for the
coal to retain the original bit. This time is called the
information retention time. How long is it for a massive black
hole? The answer can easily be deduced from the known luminosity
of black holes.  In $(3+1)$ dimensions one finds
\begin{eqnarray}
t_R \sim G^2M^3
\end{eqnarray}
For times much shorter than $t_R $ we can expect that information
which has been absorbed by the thermal horizon to be inaccessible.

\bigskip

{\it{Quantum Xerox Censorship}} \\ Let us return to the thought
experiment in figure (9) designed to detect information
duplication behind the horizon. The resolution of the dilemma  is
as follows. The point (c) must occur before  the trajectory of $O$
intersects the singularity. On the other hand $O$ may not cross
the horizon until the information retention time has elapsed. The
implication of these two constraints is most easily seen using KS
coordinates
\begin{eqnarray}
X^+&=&\rho e^{\omega} \cr
X^-&=&-\rho e^{-\omega}\cr
\omega &=&{t
\over {4MG}}
\end{eqnarray}
An observer outside the horizon must wait a time $t \sim M^3G^2$
to collect a bit from the Hawking radiation. Thus she may not jump
into the black hole until $(X^+ \sim e^{M^2G})$. On the other hand
the singularity is at $X^+X^- = M^2G^2$. This means that $O$ will
hit the singularity at a point satisfying
\begin{equation}
X^-< \exp{-M^2G}
\end{equation}
Thus for the original infalling system to send a signal which will
reach $O$ before she hits the singularity, the message must be
sent within a time interval $\delta t$  of the same order of
magnitude, an incredibly short time.

Classically there is no obstruction to sending as much information
as you like in as small a time as you like using as little energy
as you like. Quantum physics changes this. A bit of information
requires at least one quantum to transmit it. The uncertainty
principle requires that the quantum have energy of order $(\delta
t)^{-1}$. Thus the message requires a photon of energy
\begin{equation}
E_{signal} \sim \exp{M^2 G}
\end{equation}
This is completely inconsistent with the assumption that the
entire black hole, including the q-bit had energy $M$. If the
observer at (a) had that much energy available, the black hole
would have been much heavier and its horizon would have been at a
very different place. Thus we see that quantum mechanics and
gravity conspire to prevent $O$ from detecting duplicate
information.

We can now see that there is something wrong with the usual ideas
of local quantum field theory in black hole backgrounds. The
points (a) and (b) can be  widely separated by a large space-like
separation. Quantum field theory would say that the fields at
these two points are independent commuting variables and it would
predict correlations between them. But as we have seen, these
correlations are unmeasurable by any real observer subject to the
usual limitations of relativity and quantum mechanics. If you
share the belief that a theory should not predict things which are
in principle unobservable then you must conclude that local
quantum field theory in a black hole background is the wrong
starting point.

\bigskip

{\it{Baryon Violation and Black Hole Horizons}} \\ It is generally
conceded that there are no  additive conserved quantities in a
consistent quantum theory that includes gravity except for those
that couple to long range fields. If nothing else, black hole
evaporation will lead to violations of global conservation laws
such as baryon conservation. An interesting question  is where in
the black hole geometry does the violation take place? Does it
happen at or near the almost flat horizon or only at the violently
curved singularity \cite{15}  or, is it more subtle as suggested
by black hole complementarity \cite{13}?

For definiteness lets assume that baryon violation takes place in
a conventional Grand Unification scheme such as $SU(5)$. Begin
with a system of baryons and an observer all falling freely
through the horizon of a very large black hole. Since the near
horizon limit is nearly flat it is certain that the freely falling
observer will detect negligible baryon violating effects in this
region. However as time elapses the system will enter regions in
which the curvature becomes of order the GUT scale. At that point
there is every reason to think that baryon violating effects will
be observed if the observer is in any shape to observe them.

The observer outside the black hole has a very different story to
tell. According to him, the baryonic system entered the near
horizon region where it was subjected to increasing proper
temperature. When the temperature becomes of order $M_{gut}$ the
baryons are exposed to a flux of high energy particles in the
thermal atmosphere and baryon violating processes must occur. Who
is correct?

In order to answer this question consider the propagation of a
quark through empty space. Virtual baryon violating processes of
the kind shown in the Feynman Diagram  in figure (10) are
continuously taking place. In other words the quark spends part of
the time in a virtual state with the wrong baryon number even in
empty flat space. What percentage of the time is the baryon number
wrong? One might think the answer is incredibly small given the
stability of the proton. But it is not. An explicit calculation
gives a probability of order $g^2$ where $g$ is the gauge coupling
constant. Thus the quark has the wrong baryon number about 1
percent of the time. The reason we don't see this as baryon
violation is that the lifetime of the intermediate states is of
order the gut scale. The baryon number is constantly undergoing
very rapid quantum fluctuations. The usual approximately conserved
quantum number  is the time averaged baryon number normalized to 1
for the nucleon. Now consider a quark falling through the horizon
as in figure (11). It is evident from the figure that there is a
significant probability that when the quark passes the horizon at
$t=\infty$ it has the wrong baryon number. From the viewpoint of
the infalling observer doing ordinary low energy experiments on
the baryon the fluctuation is too fast to see. However, from the
outside the rapid fluctuations slow down and the quark is caught
frozen with the wrong baryon number. Of course the this
description fails to take gravitation into account but it
nevertheless shows that understanding the apparent contradictory
descriptions involves analyzing the behavior of matter at
extremely short time scales and high frequencies.

Another thought experiment can illuminate the interplay between
gravity and quantum mechanics. Suppose an observer $O$ falls
through the horizon just before the baryon as in figure (12). This
observer sends out a signal  (photon) which interacts with the
infalling baryon and measures its baryon number. The signal is
then received by a distant observer. Let us suppose that the
experiment is arranged so that the signal-photon encounters the
baryon in a region where the temperature is at least $M_{gut}$. In
the rest frame of the infalling quark, it has a time of order
$M_{gut}^{-1}$ before it crosses the horizon. Thus the photon must
be concentrated in a wave packet of size less that or equal to
$M_{gut}^{-1}$. Its energy must be so high that it will resolve
the baryon violating virtual state and will therefore have a
finite probability of reporting baryon violation at the horizon.
Complementarity works!

\bigskip

{\it{String Theory at High Frequency}} \\ Ordinary quantum field
theory can not resolve the paradoxes of black holes. We have
already seen that Q.F.T. drastically overestimates the number of
ultraviolet degrees of freedom in the near horizon region and
leads to a divergent entropy in the thermal atmosphere. String
theory is widely believed to be a consistent quantum mechanical
framework that includes gravitation. If so it must differ from
Q.F.T. in very non-trivial ways at short times.

Although we are far from achieving a definitive understanding of
black hole complementarity in string theory, there are some simple
and suggestive ways to see that string theory is very different
from Q.F.T. at high frequency \cite{16}.

Let us consider a string falling through a horizon. For our
purposes we can approximate the horizon by the light-like surface
$X^- =0$. To study the string as it falls we use light cone
coordinates. It is conventional to use $X^+$ for the light cone
time variable. We are going to be unconventional and use $X^-$.
Thus we choose the string theory gauge
\begin{equation}
\tau = X^-
\end{equation}
The string starts out at negative $X^-$ and reaches the horizon at
$X^-=0$.

Suppose the string falls through the horizon near $X^+ = 1$. Using
\begin{eqnarray}
X^- &=& -\rho e^{-\omega} \cr X^+ &=& \rho e^{\omega}  =1
\end{eqnarray}
we find that near the string
\begin{equation}
X^- =\exp{(-2\omega)} =-\exp{(-t/2MG)}
\end{equation}

The unusual properties of strings can already be seen at the level
of free string theory. In light cone gauge a free string is
described by a set of transverse coordinates $x^m(\sigma)$ where
$0 \leq \sigma < 2 \pi$. The coordinates are expressed in terms of
harmonic oscillator variables $\alpha(n)$ and $\tilde{\alpha}(n)$.
In string units
\begin{equation}
x(\sigma) = x_{cm} + \sum_n {\alpha (n)\over n}
e^{in(\tau-\sigma)}
          + {\tilde{\alpha} (n)\over n} e^{in(\tau+\sigma)}
\end{equation}

The question that will interest us has to do with the spatial size
of the string. For simplicity we will consider the ground state of
the string which classically has zero size. We usually envision
the quantum fluctuations to spread the string over a size of order
$l_s$, the string scale. However explicit calculation gives a very
different result. The spatial size $R$ will be defined in an
obvious way.
\begin{equation}
R^2=\langle 0| (x-x_{cm})^2 |0 \rangle
\end{equation}

Using the standard commutation rules for the $\alpha's$ we find
\begin{equation}
R^2= \sum_n  {1\over n} = \log(\infty)
\end{equation}
Evidently the spatial size of the string is dependent of the
frequency cutoff. If the frequency cutoff for a given observation
is $n_{max}$ then the apparent size of the string is
\begin{equation}
R^2=\log n_{max}
\end{equation}

We see a small string only if we average over sufficient time
($\tau$) to eliminate the very high frequencies. This lesson is an
important one and it will be repeated later in the form of the
{\it{ultraviolet infrared connection}} in lecture III.

Consider the outside observer's description of the infalling
string as it approaches the horizon. At any given point  the
string has a light cone time $|\tau|$ before it crosses the
horizon at $\tau=0$. Thus it makes no sense for the outside
observer to average modes of frequency smaller than $|\tau|^{-1}$.
In other words the frequency cutoff appropriate for an outside
observer increases as the horizon is approached. Using eq.(1.47)
and setting $n_{max} = |\tau|^{-1}$ we find
\begin{equation}
R^2= \log{\tau} = t/{2MG}
\end{equation}
Free string theory predicts that as a string falls toward the
horizon it grows and appears to become an increasing tangled mass
of string but only to the external observer. The infalling
observer, depending on how she interacts with the string has a
fixed time resolution and sees no growth.

\bigskip

{\it{The Space Time Uncertainty Relation}} \\ Even more revealing are
the fluctuations of the longitudinal \cite{17} coordinate  $X^+$
(usually called $X^-$). First consider a classical point particle.
It crosses the horizon, ($X^-=0$), at a finite value of $X^+$. At
that point the radial space-like distance from the horizon
vanishes.
\begin{equation}
\rho^2 = -X^+X^- =0
\end{equation}

Now consider the falling string. The coordinate $X^+(\sigma)$ is
not an independent variable in string theory. To find out how it
behaves we use the constraint equation
\begin{equation}
\partial_{\sigma}X^+ =\partial_{\sigma}x^i \partial_{\tau}x^i
\end{equation}
The fact that the string does not require an independent degree of
freedom for fluctuations in the $X^-$ direction was one of the
early indications of the large reduction in the number of degrees
of freedom expected in a holographic theory. Using eq.(1.50) we
can express $X^+(\sigma) $ in terms of harmonic oscillators. An
explicit calculation gives
\begin{eqnarray}
(\Delta X^+)^2 &\equiv& \langle 0|(X^+ -X^+_{cm})^2 |0 \rangle \cr
&=& l_s^2 \sum_n  {1 \over n^3} \cr &=& l_s^2 n_{max}^2
\end{eqnarray}

This is a special case of a fundamental new uncertainty relation
\cite{17} \cite{18} which occurs throughout string theory and
which we will return to. To write it in a more suggestive form we
write $n_{max} = (\Delta \tau)^{-1}$ or equivalently $n_{max} =
(\delta X^-)^{-1}$. Equation (1.51) then takes the symmetrical
form
\begin{equation}
\Delta X^+ \Delta X^- = l_s^2
\end{equation}
This is the {\it{string uncertainty principle}}. It implies that
there is a fundamental unit of area in the $X^+,X^-$ plane. It is
reminiscent of uncertainty principles which occur in
non-commutative geometry but it is not put in by hand.

To appreciate the implications of the space time uncertainty
relation, let us consider an infalling massless string whose
center of mass moves along the trajectory $X^+ = 1$. As $X^-$
tends to zero the fluctuation in $X^+$, as seen by an outside
observer, increases like $l_s^2 /X^-$. Thus  the stringy matter
will be  spread over  region $X^+ X^- \leq l_s^2$. From the point
of view of Schwarzschild coordinates, instead of asymptotically
approaching the horizon, the stringy matter can not be localized
more precisely than to say that it is within a proper distance
$l_s$ from the horizon.

What we are seeing is a new relativity principle. According to the
usual relativity principles, two observers in relative motion will
disagree about the length of rods and the rate of clocks. But
there is an invariant concept, the event, which occurs at a well
defined space-time location. Even this is eliminated by black hole
complementarity. External and freely falling observers will
radically disagree about where and when events such as baryon
violation take place or where the energy and momentum of a string
is located. As we have seen, quantum mechanics and relativity
conspire to insure that no observer ever sees a violation of the
laws of quantum mechanics.

We have also seen that the origin of this relativity of
descriptions is the behavior of the very high frequency
fluctuations which are invisible to the freely falling observer
but which dominate the description of the outside observer.

How can it be that the usual ideas of local quantum field theory
fail so badly? In the remaining lectures we will see that
conventional ideas of locality badly overestimate the number of
independent degrees of freedom of a system. The key to black hole
complementarity is the vast reduction implied by the holographic
principle.

\setcounter{equation}{0}
\section{Entropy Bounds}

\bigskip

{\it{Maximum Entropy}}

The Holographic Principle is about the counting of quantum states
of a system. We begin by considering a large region of space
$\Gamma$. For simplicity we take the region to be a sphere. Now
consider the space of states that describe arbitrary systems that
can fit into $\Gamma$ such that the region outside $\Gamma$ is
empty space. Our goal is to determine the dimensionality of that
state-space. Lets consider some preliminary examples. Suppose we
are dealing with a lattice of spins. Let the lattice spacing be
$a$ and the volume of $\Gamma$ be $V$. The number of spins is
$V/a^3$ and the number of orthogonal states supported in $\Gamma$
is
\begin{equation}
N_{states}=2^{V \over {a^3}}
\end{equation}

A second example is a continuum quantum field theory. In this case
the number of quantum states will diverge for obvious reasons. We
can limit the states, for example by requiring the energy density
to be no larger than some bound $\rho_{max}$. In this case the
states can be counted using some concepts from thermodynamics. One
begins by computing the thermodynamic entropy density $s$ as a
function of the energy density $\rho$. The total entropy is
\begin{equation}
S=s(\rho)V
\end{equation}
The total number of states is of order
\begin{equation}
N_{states} \sim \exp{S}= \exp s(\rho_{max})V
\end{equation}
In each case the number of distinct states is exponential in the
volume $V$. This is a very general property of conventional local
systems and represents the fact that the number of independent
degrees of freedom is additive in the volume.

In counting the states of a system the entropy plays a central
role. In general entropy is not really a property of a given
system but also involves ones state of knowledge of the system. To
define entropy we begin with some restrictions that express what
we know, for example, the energy within certain limits, the
angular momentum and whatever else we may know. The entropy is by
definition the logarithm of the number of quantum states that
satisfy the given restrictions.

There is another concept that we will call the {\it{maximum
entropy}}. This is a property of the system. It is the logarithm
of the total number of states. In other words it is the entropy
given that we know nothing about the state of the system. For the
spin system the maximum entropy is
\begin{equation}
S_{max} = {V \over a^3} \log{2}
\end{equation}
This is typical of the maximum entropy. Whenever it exists it is
proportional to the volume. More precisely it is proportional to
the number of simple degrees of freedom that it takes to describe
the system.

Let us now consider a system that includes gravity. Again we focus
on a spherical region of space $\Gamma$ with a boundary $\partial
\Gamma $. The area of the boundary is $A$. Suppose we have a
thermodynamic system with entropy $S$ that is completely contained
within $\Gamma$. The total mass of this system can not exceed the
mass of a black hole of area $A$ or else it will be bigger than
the region.

Now imagine collapsing a spherically symmetric shell of matter
with just the right amount of energy so that together with the
original mass it forms a black hole which just fills the region.
In other words the area of the horizon of the black hole is $A$.
This is shown in figure (13). The result of this process is a
system of known entropy, $S = A/4G$. But now we can use the second
law of thermodynamics to tell us that the original entropy inside
$\Gamma$ had to be less than or equal to $A/4G$. In other words
the maximum entropy of a region of space is proportional to its
area measured in Planck units.   Such bounds are called
{\it{holographic}}.

\bigskip

{\it{Entropy on Light-Like Surfaces}}

We will see that it is most natural to define holographic entropy
bounds on light-like surfaces \cite{2} as opposed to space-like
surfaces. Under certain circumstances the bounds can be translated
to space-like surfaces but not always.

Let us start with an example in asymptotically flat space-time. We
assume that flat Minkowski coordinates $X^+,X^-, x^i$ can be
defined at asymptotic distances. In this lecture we will revert to
the usual convention in which $X^+$ is used as a light cone time
variable. We will now define a "light sheet". Consider the set of
all light rays which lie in the surface $X^+ = X^+_0$ in the limit
$X^- \to +\infty$. In ordinary flat space this congruence of rays
define a flat 3-dimensional  light-like surface. In general they
define a light like surface called a light sheet. The light sheet
will typically have singular caustic lines but can be defined in a
unique way \cite{19}. When we vary $X^+_0$ the light sheets fill
all space-time except for those points that lie behind black hole
horizons.

Now consider a space-time point $p$. We will assign it light-cone
coordinates as follows. If it lies on the light sheet $X^+_0$ we
assign it the value $X^+=X^+_0$. Also if it lies on the light ray
which asymptotically has transverse coordinate $x^i_0$ we assign
it $x^i=x^i_0$. The value of $X^-$ that we assign will not matter.
The  two dimensional $x^i$ plane is called the Screen.

Next assume  a black hole  passes through the light sheet $X^+_0$.
The stretched  horizon \footnote{The stretched horizon is a
time-like surface just outside the mathematical light-like
surface. Its precise definition is not important here.} of the
black hole describes a two dimensional surface in the 3
dimensional light sheet as shown in figure (14). Each point on the
stretched horizon has unique coordinates $X^+,x^i$. More generally
if there are several black holes passing through the light sheet
we can map each of their stretched horizons to screen in a single
valued manner.

Since the entropy of the black hole is equal to   $1/4G$ times the
area of the horizon we can define an entropy density of $1/4G$ on
the stretched horizon. The mapping to the screen then defines an
entropy density in the $x^i$ plane, $\sigma(x)$. It is a
remarkable fact that $\sigma(x)$ is always less than or equal to
$1/4G$.

To prove that $\sigma(x) \leq {1 \over 4G}$ we make use of the
focusing theorem of general relativity. The focusing theorem
depends on the positivity of energy and is based on the tendency
for light to bend around regions of non-zero energy.  Consider
bundle of light rays with cross sectional area $\alpha$. The light
rays are parameterized by an affine parameter $\lambda$. The
focusing theorem says that
\begin{equation}
{d^2 \alpha \over d \lambda ^2 } \leq 0
\end{equation}

Consider a bundle of light rays in the light sheet which begin on
the stretched horizon and go off to $X^-=\infty$. Since the light
rays defining the light sheet are parallel in the asymptotic
region $d\alpha/{d\lambda} \to 0$. The focusing theorem tells us
that as we work back toward the horizon, the area of the bundle
decreases. It follows that the image of a patch of horizon on the
screen is larger than the patch itself. The holographic bound
immediately follows.
\begin{equation}
\sigma(x) \leq {1 \over 4G}
\end{equation}

This is a surprising conclusion. No matter how we distribute the
black holes in 3 dimensional space, the image of the entropy on
the screen  always satisfies the entropy bound (2.6). An example
which helps clarify how this happens involves two black holes.
Suppose we try to hide one of them behind the other along the
$X^-$ axis, thus doubling the entropy density in the $x$ plane.
The bending and focusing of light always acts as in figure (15) to
prevent $\sigma (x)$ from exceeding the bound. These
considerations lead us to the more general conjecture that for any
system, when it is mapped to the screen the entropy density obeys
the bound (2.6).

\bigskip

{\it{Robertson Walker Geometry}}

This kind of bound has been generalized
to {\it{flat}} Robertson Walker geometries by Fischler and
Susskind \cite{20} and to more general geometries by Bousso
\cite{21} \cite{22}. First review the RW case. We will consider
the general case of $d+1$ dimensions. The metric has the form
\begin{equation}
ds^2=dt^2 -a(t)^2 dx^mdx^m
\end{equation}
where the index $m$ runs over  the $d$ spatial directions. The
function $a(t)$ is assumed to grow as a power of $t$.
\begin{equation}
a(t) = a_0 t^p
\end{equation}
Lets also make the usual simplifying cosmological assumptions of
homogeneity. In particular we assume that the spatial entropy
density (per unit $d$ volume) is homogeneous. Later, following
Bousso, we will relax these assumptions.

At time $t$ we consider a spherical region $\Gamma$ of volume $V$
and area $A$. The boundary $(d-1)$-sphere, $\partial \Gamma$, will
play the same role as the screen in the previous discussion. The
light-sheet is now defined by the backward light cone formed by
light rays that propagate from $\partial \Gamma$ into the past.

As in the previous  case the holographic bound applies to the
entropy passing through the light sheet. The bound states that the
total entropy passing through the light sheet does not exceed
$A/4G$. The key to a proof is again the focusing theorem. We
observe that at the screen the area of the outgoing bundle of
light rays is increasing as we go to later times. In other words
the light sheet has positive expansion into the future and
negative expansion into the past. The focusing theorem again tells
us that if we map the entropy of black holes passing through the
light sheet to the screen, the resulting density satisfies the
holographic bound.

It is now easy to see why we concentrate on light sheets instead
of space like surfaces. Obviously if the spatial entropy density
is uniform and we choose $\Gamma$ big enough, the entropy will
exceed the area. However if $\Gamma$ is larger than the particle
horizon at time $t$ the light sheet is not a cone but rather a
truncated cone which is cut off by the big bang at $t=0$. Thus a
portion of the entropy present at time $t$ never passed through
the light sheet. If we only count that portion of the entropy
which did pass through the light sheet it will scale like the area
$A$. We will return to the question of space-like bounds after
discussing Bousso's generalization \cite{21} of the FS bound.

\bigskip

{\it{Bousso's Generalization}}

Consider an arbitrary cosmology. Take a space-like region $\Gamma$
bounded by the space-like boundary $\partial \Gamma$. Following
Bousso \cite{21}, at any point on the boundary we can construct
four light rays that are perpendicular to the boundary. We will
call these the four {\it{branches}}. Two branches go toward the
future. One of them is composed of  outgoing rays and the other is
ingoing. Similarly two branches  go to the past. On any   of these
branches a light ray, together with its neighbors define a
positive or negative expansion as we move away from the boundary.
In ordinary flat space-time if $\partial \Gamma$ is convex the
outgoing (ingoing) rays have positive (negative) expansion.
However in non-static universes other combinations are possible.
For example in a rapidly contracting universe the outgoing  future
branch may have negative expansion.

If we consider general boundaries the sign of the expansion of a
given branch may vary as we move over the surface. For simplicity
we restrict attention to those regions for which a given branch
has a unique sign. We can now state Bousso's rule:

From the boundary $\partial \Gamma$ construct all light sheets
which have negative expansion as we move away. These light sheets
may terminate at the tip of a cone or a caustic or even a boundary
of the geometry. Bousso's bound states that the entropy passing
through these light sheets is less that $A/4G$ where $A$ is the
boundary of $\partial \Gamma$.

To help visualize how Bousso's construction works we will consider
spherically symmetric geometries and use Penrose diagrams to
describe them. The Penrose diagram represents the radial and time
directions. Each point of such a diagram really stands for a
2-sphere (more generally a ($d-1$)-sphere). The four branches at a
given point on the Penrose diagram are represented by a pair of 45
degree lines passing through that point.  However we are only
interested in the branches of negative expansion. For example in
figure(16) we illustrate flat space-time and the negative
expansion branches of a typical local 2-sphere.

In general as we move around in the Penrose diagram the particular
branches which have negative expansion may change. For example if
the cosmology initially expands and then collapses, the outgoing
future branch will go from positive to negative expansion. Bousso
introduced a notation to indicate this. The Penrose diagram is
divided into a number of regions depending on which branches have
negative expansion. In each region the negative expansion branches
are indicated by their directions at a typical point. Thus in
figure(17) we draw the Penrose- Bousso  (PB) diagram for a
positive curvature, matter dominated universe that begins with a
bang and ends with a crunch. It consists of four distinct regions.

In region I of figure (17) the expansion of the universe causes
both past branches to have negative expansion. Thus we draw light
surfaces into the past. These light surfaces terminate on the
initial boundary of the geometry and are similar to the truncated
cones that we discussed in the flat RW case. The holographic bound
in this case says that the entropy passing through either backward
light surface is bounded by the area of the 2-sphere at point $p$.
Bousso's rule tells us nothing in this case about the entropy on
space like surfaces bounded by $p$.

Now move on to region II. The relevant light sheets in this region
begin on the 2-sphere $q$ and both terminate at the spatial
origin. These are untruncated cones and the entropy on both of
them is holographically bounded. There is something new in this
case. We find that the entropy is bounded on a future light sheet.
Now consider a space like surface bounded by $q$ and extending to
the spatial origin. It is evident that  any matter which passes
through the space-like surface must also pass through the future
light sheet. By the second law of thermodynamics the entropy on
the space-like surface can not exceed the entropy on the future
light sheet. Thus in this case the entropy in a space-like region
can be holographically bounded. Thus, one condition for a
space-like bound is that the entropy is bounded by a corresponding
future light sheet. With this in mind we return to region I. For
region I there is no future bound and therefore the entropy is not
bounded on space-like regions with boundary $p$.

In region III the entropy bounds are both on future light sheets.
Nevertheless there is no space-like bound. The reason is that not
all  matter which passes through space-like surfaces is  forced to
pass through the future light sheets.

Region IV is identical to region II with the spatial origin being
replaced by the diametrically opposed antipode. Thus we see that
there are light-like bounds in all four regions but only in II and
IV are there holographic bounds on space-like regions.

Another example of interest is inflationary cosmology. The PB
diagram for de-Sitter space  is shown in figure (18a). This time
region I has both light sheets pointing to the future. This is due
to the fact that de-Sitter space is initially contracting. In
order to describe inflationary cosmology we must terminate the de
sitter space at some late time and attach it to a conventional RW
space. This is shown in figure (18b). The dotted line where the
two geometries are joined is the reheating surface where the
entropy of the universe is created.

Let us focus on the point $p$ in figure (18b). It is easy to see
that in an ordinary inflationary cosmology $p$ can be chosen so
that the entropy on the space-like surface $p-q$ is bigger than
the area of $p$. However Bousso's rule applied to point (p) only
bounds the entropy on the past light sheet. In this case most of
the newly formed entropy on the reheating surface is not counted
since it never passed through the past light sheet. Typical
inflationary cosmologies can be studied to see that the past light
sheet bound is not violated.

As a final example we consider anti-de Sitter (AdS) space. The PB
diagram consists of an infinite strip bounded on the left by the
spatial origin and of the right by the AdS boundary. The PB
diagram consists of a single region in which both negative
expansion light sheets point toward the origin. Let us consider a
static surface of large area $A$ far from the spatial origin. The
surface is denoted by the dotted vertical line $L$ in figure (19).
We will think of $L$ as an infrared cutoff.

Consider an arbitrary point $p$ on $L$. Evidently Bousso's rules
bound the entropy on past and future light sheets bounded by $p$.
Therefore the entropy on any space-like surface bounded by $p$ and
including the origin is also holographically bounded. In other
words the entire region to the left of $L $ can be foliated with
space-like surfaces such that the maximum entropy on each surface
is $A/4G$.

AdS space is an example of a special class of geometries which
have time-like killing vectors and which can be foliated by
surfaces that satisfy the Holographic bound. These two properties
imply a very far reaching conclusion. All physics taking place in
such backgrounds (in the interior of the infrared cutoff $L$) must
be described in terms of a Hamiltonian that acts in a Hilbert
space of dimensionality
\begin{eqnarray}
N_{states} = \exp (A/4G)
\end{eqnarray}
The holographic description of AdS space is the subject of the
next lecture.


\setcounter{equation}{0}
\section{The AdS/CFT Correspondence and the Holographic Principle}

\bigskip

{\it{AdS Space}}

As we saw in Lecture II, AdS space enjoys certain properties which
make it a natural candidate for a holographic Hamiltonian
description. In this lecture we will review the holographic
description of $AdS(5) \otimes  S(5)$ \cite{23}
\cite{24}\cite{25}. Maldacena, in his lectures to this school has
explained how this space arises in type $IIb$ string theory,
either as the near horizon geometry of a stack of D3-branes  or as
a solution of ten dimensional supergravity. We will begin with a
brief review of AdS geometry.

For our purposes $5$ dimensional AdS space may be considered to be
a solid $4$ dimensional spatial ball times  the infinite time
axis. The geometry can be described by dimensionless coordinates
$t, r, \Omega$ where $t$ is time, $r$ is the radial coordinate $(0
\leq r < 1)$ and $\Omega$ parametrizes the unit $3$-sphere. The
geometry has uniform curvature  $R^{-2}$ where $R$ is the radius
of curvature. The metric we will use is
\begin{equation}
ds^2 = {R^2 \over (1-r^2)^2} \left \{(1+r^2)^2dt^2 -4dr^2 -4r^2
d\Omega^2 \right \}
\end{equation}

There is another form of the metric which is in common use,
\begin{equation}
ds^2 ={R^2 \over y^2} \left \{ dt^2  -dx^i dx^i -dy^2 \right \}
\end{equation}
where $i$ runs from $1$ to $3$.

The metric (3.2) is related to (3.1) in two different ways. First
of all it is an approximation to (3.1) in the vicinity of a point
on the boundary at $r=1$.  The $3$ sphere is replaced by the flat
tangent plane parameterized by $x^i$ and the radial coordinate is
replaced by $y$ with  $y = (1-r)$.

The second way that (3.1) and (3.2) are related is that (3.2) is
the exact metric of an incomplete patch of AdS space. A time-like
geodesic can get to $y= \infty$ in a finite proper time so that
the space in eq. (3.2) is not geodesically complete. As discussed
in the lectures of Maldacena the metric (3.2)  describes the near
horizon geometry of a stack of D3-branes located at the horizon
$y=\infty$. The metric (3.2) may be expressed in terms of the
coordinate $z=1/y$.
\begin{equation}
ds^2 =R^2  \left \{ z^2(dt^2  -dx^i dx^i) -{1\over z^2}dz^2 \right
\}
\end{equation}
In this form the horizon is at $z=0$ and the boundary is at
$z=\infty$.

To construct the space $AdS(5) \otimes S(5)$ all we have to do is
define 5 more coordinates $\omega_5$ describing the unit 5 sphere
and add a term to the metric
\begin{equation}
ds_5^2=R^2 d \omega_5^2
\end{equation}

Although the boundary of AdS is an infinite proper distance from
any point in the interior of the ball, light can travel to the
boundary and back in a finite time. For example, it takes a total
amount of (dimensionless ) time $t=\pi$ for light to make a round
trip  from the origin at $r=0$ to the boundary at $r=1$ and back.
For all practical purposes AdS space behaves like a finite cavity
with reflecting walls. The size of the cavity is of order $R$. In
what follows we will think of the cavity size $R$ as being much
larger than any microscopic scale such as the Planck or string
scale.

\bigskip

{\it{Holography in AdS Space}}

In order to have a benchmark for the counting of degrees of
freedom in $AdS(5)\otimes S(5)$  imagine constructing a cutoff
field theory in the interior of the ball. A conventional cutoff
would involve a microscopic length scale such as the 10
dimensional Planck length $l_p$. One way to do this  would be to
introduce a spatial lattice in  nine dimensional space . It is not
generally possible to make a regular lattice but a random lattice
with an average spacing $l_p$ is possible. We can then define a
simple theory such as a   Hamiltonian lattice theory on the space.
In order to count degrees of freedom we also need to regulate area
of the boundary of AdS which is infinite. The way to do that was
hinted at in lecture II. We introduce a  surface $L$ at
$r=1-\delta$. The total 9 dimensional spatial volume in the
interior of $L$ is easily computed using the metric (3.1).
\begin{equation}
V(\delta) \sim {R^9 \over {\delta}^3 }
\end{equation}
and the number of lattice sites and therefore the number of
degrees of freedom is
\begin{equation}
{V \over {l_p^9}} \sim {1\over \delta^3}{R^9 \over l_p^9}
\end{equation}
In such a theory we also will find that the maximum entropy is of
the same order of magnitude.

On the other hand the holographic bound discussed in lecture II
requires the maximum entropy and the number of degrees of freedom
to be of order
\begin{equation}
S_{max }\sim{A \over l_p^8}
\end{equation}
where $A$ is the 8 dimensional area of the boundary $L$. This is
also easily computed. We find
\begin{equation}
S_{max} \sim {1 \over \delta^3}{R^8 \over l_p^8}
\end{equation}
In other words  when $R/l_p$ becomes large the holographic
description requires a  reduction  in the number of independent
degrees of freedom by a factor $l_p/R$. To say it slightly
differently, the holographic principle implies a complete
description of all physics in the bulk of a very large AdS space
in terms of only $l_p/R$ degrees of freedom per spatial Planck
volume.

\bigskip

{\it{The AdS/CFT Correspondence}}

The correspondence between string theory in $AdS(5) \otimes S(5)$
and  Super Yang Mills (SYM) theory on the boundary has been
discussed in other lectures in this school and we will only review
some of the salient features. The correspondence states that there
is a complete equivalence between superstring theory in the bulk
of $AdS(5) \otimes S(5)$ and maximally supersymmetric  (16 real
supercharges), $3+1$ dimensional, $SU(N)$, SYM theory on the
boundary of the AdS space \cite{23}\cite{24}\cite{25}. In these
lectures SYM theory will always refer to this particular version
of supersymmetric gauge theory.

It is well known that SYM is conformally invariant and therefore
has no dimensional parameters. It will be convenient to define the
theory to live on the boundary parametrized by the dimensionless
coordinates $t,\Omega$ or $t,x$. The corresponding momenta are
also dimensionless. In fact we will use the convention that all
SYM quantities are dimensionless. On the other hand the bulk
gravity theory quantities such as mass, length and temperature
carry their usual dimensions. To convert from SYM to bulk
variables the conversion factor is $R$. Thus if $E_{sym}$ and $M$
represent the energy in the SYM and bulk theories

$$E_{sym} = MR $$
\bigskip
Similarly bulk time intervals are given by multiplying the $t$
interval by $R$.

One might think that the boundary of $AdS(5) \otimes S(5)$ is
$(8+1)$ dimensional but there is an  important sense in which it
is $3+1$ dimensional. To see this let us Weyl rescale the metric
by a factor ${R^2 \over (1-r^2)^2}$ so that the rescaled metric at
the boundary is finite. The new metric is
\begin{equation}
dS^2 =  \left \{(1+r^2)^2dt^2 -4dr^2 -4r^2 d\Omega^2 \right \}
      + \left \{(1-r^2)^2 d\omega_5^2 \right \}
\end{equation}
Notice that the size of the 5-sphere shrinks to zero  as the
boundary  at $r=1$ is approached. The boundary of the geometry is
therefore $3+1$ dimensional.

Let us return to the correspondence between the bulk and bounday
theories. The ten dimensional bulk theory has two dimensionless
parameters. These are:

1. The radius of curvature of the AdS space measured in string
units $R/l_s$

2. The dimensionless string coupling constant $g$.

The string coupling constant and length scale are are related to
the ten dimensional Planck length and Newton constant by
\begin{equation}
l_p^8=g^2 l_s^8 = G
\end{equation}

On the other side of the correspondence, the gauge theory also has
two constants. They are

1. The rank of the gauge group $N$

2. The gauge coupling $g_{ym}$

The relation between the string and gauge parameters was given by
Maldacena \cite{23}. It is
\begin{eqnarray}
{R \over l_s} &=& ({Ng_{ym}^2})^{1\over 4} \cr g&=&g_{ym}^2
\end{eqnarray}
We can also write ten dimensional Newton constant in the form
\begin{eqnarray}
G=R^8 /N^2
\end{eqnarray}

There are two distinct limits that are especially interesting,
depending on one's motivation. The AdS/CFT correspondence has been
widely studied as a tool for learning about the behavior of gauge
theories in the strongly coupled 't Hooft limit. From the gauge
theory point of view the 't Hooft is defined by
\begin{eqnarray}
g_{ym} & \to & 0 \cr N &\to& \infty \cr g_{ym}^2 N &=& constant
\end{eqnarray}
From the bulk string point of view the limit is
\begin{eqnarray}
g &\to& 0 \cr {R\over l_s} &=& constant
\end{eqnarray}
Thus the strongly coupled 't Hooft limit is also the classical
string theory limit in a fixed and large AdS space. This limit is
dominated by classical supergravity theory.

The interesting limit from the viewpoint of the holographic
principle is a different one. We will be interested in the
behavior of the theory as the AdS radius increases but with the
parameters that govern the microscopic physics   in the bulk kept
fixed. This means we want the limit
\begin{eqnarray}
g&=& constant \cr R/l_s & \to & \infty
\end{eqnarray}
On the gauge theory side this is
\begin{eqnarray}
g_{ym}&=& constant \cr N &\to& \infty
\end{eqnarray}
Our goal will be to show that the number of quantum degrees of
freedom in the gauge theory description satisfies the holographic
behavior in eq. (3.8).

\bigskip

{\it{The Infrared Ultraviolet Connection}}

In either of the metrics (3.1) or (3.2) the proper area of any
finite coordinate patch tends to $\infty$ as the boundary of AdS
is approached. Thus we expect that the number of degrees of
freedom associated with such a patch should diverge. This is
consistent with the fact that a continuum quantum field theory
such as SYM has an infinity of modes in any finite three
dimensional patch. In order to do a more refined counting
\cite{26} we need to regulate both the area of the AdS boundary
and the number of ultraviolet degrees of freedom in the SYM. As we
will see, these apparently different regulators are really two
sides of the same coin. We have already discussed infrared (IR)
regulating the area of AdS by introducing a surrogate boundary $L$
at $r=1 - \delta$ or similarly at $y=\delta$.

That the the IR regulator of the bulk theory is equivalent to an
ultraviolet (UV) regulator in the SYM theory is called the IR/UV
connection \cite{26}. It can be motivated in a number of ways. In
this lecture we give an argument based on the quantum fluctuations
of the positions of the D3-branes which are nominally located at
the origin of the  coordinate $z$ in eq. (3.3). The location of a
point on a 3 brane  is defined by six coordinates $z, \omega_5$.
We may also choose the six coordinates to be cartesian coordinates
$(z^1, ...,z^6)$. The original coordinate $z$ is defined by
\begin{equation}
z^2 = (z^1)^2 +...+(z^6)^2
\end{equation}

The coordinates $z^m$ are represented in the SYM theory by six
scalar fields on the world volume of the branes. If the six scalar
fields  $\phi^n$ are canonically  normalized then the precise
connection between the $z's$ and $\phi's$ is
\begin{equation}
z={g_{ym}l_s^2 \over R^2} \phi
\end{equation}
Strictly speaking eq.(3.18) does not make sense because the fields
$\phi$ are $N \times N$ matrices. The situation is the same as in
matrix theory where we identify the $N$ eigenvalues of the
matrices in eq.(3.18) to be the coordinates $z^m$ of the $N$
D3-branes. As in matrix theory the geometry is noncommutative and
only configurations in which the six matrix valued fields commute
have a classical interpretation. However the radial coordinate $z=
\sqrt{z^mz^m}$ can be defined by
\begin{equation}
z^2 = \left({g_{ym}l_s^2 \over R^2} \right)^2 {1 \over N}Tr
{\phi}^2
\end{equation}

A question which is often asked is; Where are the D3-branes
located in the AdS space?  The usual answer is that they are at
the horizon $z=0$. However our experiences in lecture I with
similar questions should warn us that the answer may be  more
subtle. In lecture I ( see the discussion from eq(1.45) to eq.(
1.52) ) a  question was asked about the location of a string. What
we found is that the answer depends on what frequency range it is
probed with. High frequency or short time probes see the string
widely spread in space while low frequency probes see a well
localized string.

To answer the corresponding question about D3-branes we need to
study the quantum fluctuations of their position. The fields
$\phi$ are scalar quantum fields whose scaling dimensions are
known to be exactly $(length)^{-1}$. From this it follows that any
of the $N^2$ components of $\phi$ satisfies
\begin{equation}
\langle \phi_{ab}^2 \rangle \sim \delta^{-2}
\end{equation}
where $\delta$ is the ultraviolet regulator of the field theory.
It follows from eq(3.20) that the average value of $z$ satisfies
\begin{equation}
<z>^2 \sim \left({g_{ym}l_s^2 \over R^2} \right)^2 {N \over
\delta^2}
\end{equation}
or, using eq's(3.12)
\begin{equation}
<z>^2  \sim \delta^{-2}
\end{equation}

In terms of the coordinate $y$ which vanishes at the boundary of
AdS
\begin{equation}
<y>^2 \sim \delta^2
\end{equation}

Evidently low frequency probes see the branes at $z=0$ but as the
frequency of the probe increases the brane appears  to move toward
the boundary at $z=\infty$.  The precise connection between the UV
SYM cutoff and the bulk-theory IR cutoff is given by eq.(3.23).

\bigskip

{\it{Counting Degrees of Freedom}}

Let us now turn to the problem of counting the number of degrees
of freedom needed to describe the region $y > \delta$ \cite{26}.
The UV/IR connection implies that this region can be described in
terms of an ultraviolet regulated theory with a cutoff length
$\delta$. Consider a patch of the boundary with unit coordinate
area. Within that patch there are $1/\delta^3$ cutoff cells of
size $\delta$. Within each such  cell the fields are constant in a
cutoff theory. Thus each cell has of order $N^2$ degrees of
freedom corresponding to the $N \otimes N$ components of the
adjoint representation of $U(N)$. Thus the number of degrees of
freedom on the unit area is
\begin{equation}
N_{dof} = {N^2 \over \delta^3}
\end{equation}

On the other hand the 8-dimensional area of the regulated patch is
\begin{equation}
A={R^3 \over \delta^3}\times R^5= {R^8 \over \delta^3}
\end{equation}
and the number of degrees of freedom per unit area is
\begin{equation}
{N_{dof} \over A}\sim {N^2 \over R^8}
\end{equation}
Finally we may use eq.(3.12)
\begin{equation}
{N_{dof} \over A} \sim {1 \over G}
\end{equation}
This is exactly what is required by the holographic principle.

\bigskip

{\it{AdS Black Holes}}

The apparently irreconcilable demands of black hole thermodynamics
and the principles of quantum mechanics have led us to a very
strange view of the world as a hologram. Now we will return, full
circle, to see how the holographic description of $AdS(5) \otimes
S(5)$ provides a description of black holes. What would be most
interesting would be to give a holographic description of
10-dimensional black hole formation and evaporation in an $AdS(5)
\otimes S(5)$ space which is much larger than the black hole.
Unfortunately we will see that this is far beyond our present
ability. There are however, black hole solutions in $AdS(5)
\otimes S(5)$ which are within our current understanding. These
are the black holes which have  Schwarzschild radii as large or
larger than the radius of curvature $R$. Such black holes are
stable against decay and do not evaporate.   In fact these black
holes homogeneously fill the 5-sphere. They are solutions of the
dimensionally reduced 5-dimensional Einstein equations with a
negative cosmological constant. The thermodynamics can be derived
from the black hole solutions  by first computing the area of the
horizon and then using the Bekenstein Hawking formula .

One finds that the entropy is related to their mass by
\begin{equation}
S= c\left ( M^3R^{11}G^{-1 } \right )^{1 \over 4}
\end{equation}
Where $G$ is the ten dimensional Newton constant and $c$ is a
numerical constant. Using the thermodynamic relation $dM=TdS$ we
can compute the relation between mass and temperature.
\begin{equation}
M=c{R^{11}T^4 \over G}
\end{equation}
or in terms of dimensionless SYM quantities
\begin{eqnarray}
E_{sym} &=& c{ R^8 \over G}T_{sym}^4 \cr &=&cN^2 T_{sym}^4
\end{eqnarray}

Eq.(3.30) has a surprisingly simple interpretation. Recall that in
$3+1$ dimensions the Stephan-Boltzmann law for the energy density
of radiation is
\begin{equation}
E=T^4V
\end{equation}
where $V$ is the volume. In the present case the relevant volume
is the dimensionless  3-area of the  unit boundary sphere.
Furthermore there are $\sim N^2$ quantum fields in the $U(N)$
gauge theory so that apart from a numerical constant eq.(3.30) is
nothing but the Stephan-Bolzmann law for black body radiation.
Evidently the holographic description of the AdS black holes is a
simple as it could be; a black body thermal gas of $N^2$ species
of quanta propagating on the boundary hologram.

\bigskip

{\it{The Horizon}}

The  high frequency quantum fluctuation of the location of the
D3-branes are invisible to a low frequency probe. Roughly speaking
this is insured by the renormalization  group as applied to the
SYM description of the branes. The renormalization group is what
insures that  our bodies  are not  severely damaged by constant
exposure to high frequency vacuum fluctuations.   We are not protected in
the same way from classical fluctuations. An example is the
thermal fluctuations of fields at high temperature. All probes
sense thermal fluctuations of the brane locations. Let us return
to eq.(3.20) but now, instead of using eq.(3.21) we use the
thermal field fluctuations of $\phi$. For each of the $N^2$
components the thermal fluctuations have the form
\begin{equation}
< \phi^2> = T_{sym}^2
\end{equation}
and we find eqs.(3.22 ) and (3.23) replaced by
\begin{eqnarray}
<z>^2 &\sim& T_{sym}^2 \cr <y>^2 & \sim & T_{sym}^{-2}
\end{eqnarray}
It is clear that the thermal fluctuations will be strongly felt
out to a coordinate distance $z =T_{sym}$. In terms of $r$ the
corresponding position is
\begin{eqnarray}
1-r \sim 1/T_{sym}
\end{eqnarray}
In fact this coincides with the location of the horizon of the AdS
black hole.

A more precise definition of the horizon was given by Kabat and
Lifschytz \cite{27}. In the D-brane description the zero
temperature stack of branes can be thought of as an extreme black
brane with the horizon at $z=0$. We would like to find something
special about the corresponding point $\phi =0$ in the SYM
description. Let us displace one of the branes of the stack to a
classical location $z$. At zero temperature supersymmetry insures
the stability of this configuration. From the gauge theory point
of view we have shifted a scalar field and broken the gauge
symmetry to $U(1)\otimes U(N-1)$. The effect is to give the
"W-bosons" a mass $g \phi$. From the brane point of view we have
given a mass to the strings which extend between the displaced
brane at $z$ and the others at $z=0$. Now we see what is special
about $z=0$. If we place a brane probe at a distance from the
horizon there are massive modes of the brane. These modes become
massless at the horizon. Presumably if we went even further these
modes would become tachyonic and lead to an instability involving
the irreversible production of strings connecting the probe and
stack.

Kabat and Lifschytz \cite{27} conjecture that this is the general
feature of horizons in both the AdS/CFT theory and Matrix theory.
In the AdS case we begin with a spontaneously broken SYM at finite
temperature. It is well known that the mass of the $W$ boson is
corrected by finite temperature effects. Kabat and Lifschytz argue
that at finite temperature the tachyonic instability occurs at a
non-zero value of $\phi$. This value  corresponds to the position
of the horizon.

The string theory correspondence gives a fairly convincing picture
of the thermal effects on the $W$ mass \cite{27}. Let the probe
brane be at $z$. The thermal effects are represented by a black
hole or black brane with a horizon at $z_H$. We assume $z>z_H$.
Now the string connecting the probe to the stack is terminated at
the black hole horizon and its mass is
\begin{equation}
M=(z-z_H)/l_s
\end{equation}
As $z \to z_H$ the string becomes massless and then tachyonic.

\setcounter{equation}{0}
\section{The Flat Space Limit}

\bigskip

{\it{The Flat Space Limit}}

Gauge theory, gravity correspondences are especially
interesting because they provide  nonperturbative definitions of
some  quantum-gravity systems. The first example was matrix theory
which uses SYM theory to define 11 dimensional supergravity in the
DLCQ framework. To effectively decompactify the light cone
direction we must pass to the large $N$ limit keeping the gauge
coupling fixed.

It has also been proposed that the AdS/CFT correspondence can be
used to give a non-perturbative definition of type IIb string
theory \cite{28}. For this purpose we regard AdS space in the form
of eq.(3.1) as a finite cavity with reflecting walls. It provides
an ideal "box" for the purpose of infrared regulating a theory.
Although the actual metric distance from any point in the bulk
geometry to the boundary is infinite, it nevertheless closely
resembles an ordinary finite box of size $R$. For example the time
for light to propagate from $r=0$ to the boundary and back is
finite $\pi R$. Another indication of the finiteness of the box is
that the energy eigenvalues of a particle moving in the metric
(3.1) are discrete with the scale of energy being $1/R$.

To define the infinite volume limit we want to let $R \to \infty $
while keeping fixed the microscopic parameters of the theory such
as $g$ and $l_s$. We also want to keep fixed the energy  and
length scales in string units. Let us see what this means in terms
of SYM quantities. From eq's(3.11) we see that we must allow $N
\to \infty$ while keeping $g_{ym}$ fixed just as in matrix theory.
Furthermore the SYM energy is related to the mass $M$ by $ E_{sym}
= MR =M l_s(Ng_{ym}^2)^{1\over4}$ Accordingly, to keep $M$ fixed
we must allow $E_{sym}$ to grow like $ N^{1\over4}$ while time
intervals must scale like $t \to N^{-{1 \over4}}$ Matrix theory
also requires a scaling of energy with $N$ but it is different.
Instead of eq.(4.1) matrix theory involves energy of order $1/N$.

The next question is what quantities make sense in the limit
\begin{eqnarray}
N &\to & \infty \cr
g_{ym} &=& constant \cr
E_{sym} &\to N^{1
\over4}&
\end{eqnarray}

The answer must be that any quantity that has a well defined flat
space limit  in ten dimensional IIb string theory  should
correspond to a quantity with a good limit under (4.1). The most
obvious quantities are the spectrum and scattering matrix of
stable  particles. The only such particles are the massless
supergravity multiplet. This includes Kaluza-Klein particles with
non-zero momentum on the 5-sphere. From the point of view of the 5
dimensional AdS space these objects have non zero mass but they
are stable. The 5-dimensional  AdS mass of a particle with
momentum $k$ on the 5-sphere is
\begin{eqnarray}
M = |k|
\end{eqnarray}
or in terms of the $S(5)$ angular momentum $J$
\begin{eqnarray}
M = J/R
\end{eqnarray}
The existence and stability of these ten-dimensionally massless
particles has been established  beyond doubt from properties of
the SYM theory (See Maldacena's lectures). The existence and
properties of an S-matrix have also been studied
\cite{29}\cite{28} but much less can be rigorously established.
The idea for constructing scattering amplitudes is to use
appropriate local gauge-invariant operators in the boundary theory
as sources of the bulk particles. The particles can be aimed from
the boundary toward the origin ($r=0$) of the cavity and by
carefully controlling the sources they can be made to interact in
a small enough region that the curvature of the space is
irrelevant. All kinds of interesting phenomena could occur during
the collision. This includes the formation and evaporation of 10
dimensional black holes. You can look up the details of this kind
of construction in the papers by Polchinski and Susskind \cite{28}
\cite{29}. In this
lecture we will concentrate on a couple of the poorly understood
issues connected with the holographic description of in the
interior of AdS.

\bigskip

{\it{High Energy Gravitons Deep in the Bulk}}

The first issue has to do with the description of high energy
particles far from the boundary. Let us consider a massless
graviton emitted from the boundary with vanishing $S(5)$ momentum.
The creation operator for emitting the graviton is made out of the
energy-momentum tensor of the boundary theory by integrating
$T_{ij}$ with a test function whose frequency spectrum is
concentrated around some value $\omega$.
\begin{equation}
\omega =pR = p l_s(g_{ym}^2 N)^{1 \over4}
\end{equation}
Acting with the resulting operator creates a graviton of bulk
momentum $p$ propagating from the boundary toward the origin.

Once the particle has entered the bulk and passed the surrogate
boundary at $y=\delta$, the holographic principle requires that it
has a description in the regulated SYM theory with momentum cutoff
$1/\delta$. Let us first consider the case of low graviton
momentum by which we mean $pR = \omega < 1/\delta$. In this case
the source function  is slowly varying on the cutoff scale and the
ordinary renormalization group strategy applies. Integrating out
the modes beyond the cutoff results in a renormalized theory.
Because the SYM theory is scale invariant, the cutoff theory has
the same form as the original theory and the graviton is
description is the same as in the continuum theory.

However, the renormalization group does not apply to situations in
which the field theoretic source functions vary more rapidly than
the cutoff scale. Thus if ($p>\delta /R$) there is no guarantee
that the cutoff theory can describe the graviton correctly. The
problem is that the holographic principle demands that we be able
to describe all the physical states in the region $y>\delta$ by
states of the cutoff theory even if they contain high energy
gravitons.

To phrase the paradox differently, note that a massless particle
with momentum $p$ moving in the $y$ direction can be localized in
the $x$ plane with an uncertainty
\begin{equation}
R\Delta x \sim {1 \over p}
\end{equation}
Thus it should be possible to distinguish two such particles if
their separation $x$ is of order $1/pR$ or bigger. On the other
hand the largest momenta in the cutoff SYM theory is $1/\delta
<<pR$. How is it possible to construct such well localized objects
out of the low momentum modes of the SYM fields? We will argue
that the only possible answer is that the high energy  graviton is
created by operators that involve many SYM quanta. In other words
the effective operator which creates the high energy  graviton in
the cutoff theory must be high order in the fundamental SYM
fields.

The order can be estimated by taking the total dimensionless
energy $\omega$ of the graviton and dividing up among gauge quanta
of energy $1/\delta$.
\begin{equation}
n=\omega \delta =pR \delta
\end{equation}
To illustrate the point  consider an n-particle wave function (as
long as $n<<N$ the SYM quanta can be treated as non-identical
Boltzmann particles). As an example we choose a product wave
function
\begin{equation}
\psi(x_1,x_2,.....,x_n) = \psi(x_1 )\psi(x_2)....\psi(x_n)
\end{equation}
with
\begin{equation}
\psi(x) = \exp{-\left({|x|\over \delta}\right)}
\end{equation}

Note that wave functions of this type are composed of momenta of
order $1/\delta$ and make sense in the cutoff theory.

Suppose we have two such states which are identical except one of
them is displaced a distance $a$ in the $x$ direction. The inner
product of these states is given by
\begin{eqnarray}
\left\{ \int \psi^*(x) \psi(x-a)\right\}^n  \sim \exp {-na/\delta}
\end{eqnarray}
The function $exp {-na/\delta}$ in eq.(4.9) is narrowly peaked on
the cutoff scale if $n$ is large. In other words these states are
distinguishable when they are displaced by distance $\delta /n$
even though the largest individual momentum is only $1/\delta$.

Thus we see that  fine details can be distinguished in the coarse
grained theory but only if the gravitons and other bulk particles
are identified as an increasingly large number of gauge quanta as
the UV cutoff of the SYM is lowered and/or the momentum is
increased. This is very similar to matrix theory in which a
graviton of momentum $P_-$ is represented by a number of partons
which grow with $P_-$.

\bigskip

{\it{Kaluza Klein Modes}}

So far we have considered particles which are massless in the 5
dimensional sense. Now let us consider a graviton with
non-vanishing 5-momentum $k$. We want to hold $k$ fixed as we let
$R \to \infty$. The 5 dimensional mass is $k$. Let us also assume
$p$, the momentum in the $y$ direction is also kept fixed. The
dimensionless SYM energy of the state is
\begin{equation}
\omega = R \sqrt{k^2 + p^2}
\end{equation}

Once again it is known how to create such particles by introducing
a source at the boundary. The source in this case is a local gauge
invariant SYM operator of the form
\begin{equation}
S_n=Tr(\phi)^n
\end{equation}
This expression stands for an nth order monomial in the scalar SYM
fields $\phi$. The integer $n$ is equal to the $S(5)$ angular
momentum $kR$.
\begin{equation}
n=kR
\end{equation}
To construct a creation operator for a particle of momentum $p,k$
we integrate $S_n$ with a test function of frequency $\omega$
given in eq.(4.10).

The puzzling feature of this prescription is that it injects the
particle into the system with a local boundary operator. But a
massive particle with energy $\sqrt {k^2 +p^2}$ can never get near
the boundary. This can be seen from the motion of a massive
classical particle in AdS space. If a particle of mass $M$ moves
along the $y$ axis with total bulk energy $E =\omega /R$ then the
closest it comes to the boundary is
\begin{equation}
y^* = M/E
\end{equation}
where $y^*$ is the classical turning point of the trajectory. It
is also true that the solution of the classical wave equation for
such a particle has its largest value at this point.  For $y<y*$
the wave function quickly goes to zero. Somehow the local boundary
field $S_n$ must be creating bulk particles far from the boundary.

This behavior can be qualitatively be understood in an elementary
way from the SYM theory.  The operator $S_n$ in eq.(4.11)
describes the creation of $n$ quanta. Suppose that the SYM energy
$\omega$ is divided among the quanta so that each carries
$\omega/n$. Equivalently the quanta have wave length $n/ \omega$.
According to the UV/IR connection quanta of this wave length
correspond to bulk phenomena at $y=n/\omega$. Using eq's.(4.10 )
and (4.12) we see that this corresponds to the position $y*$. In
this way we see that the local operator constructed from $S_n$ by
projecting out given frequency  components actually corresponds to
a bulk particle at its classical turning point.

Before concluding this final lecture the are some negative
features of holographic descriptions which need to be mentioned.
These negative features become apparent when we begin to ask how
ordinary phenomena near the origin of a very large AdS space are
described in SYM theory \cite{30} \cite{31}. Suppose we have some
object which may be macroscopic in size but which is very much
smaller than the radius of curvature $R$. According to the UV/IR
connection if the object is near the origin only the longest
wavelength modes of the SYM fields should be important for their
description. On the 3-sphere this means the almost homogeneous
modes. The number of such homogeneous modes is of order $N^2$ and
these must be the degrees of freedom which describe entire physics
within a region of size $R$ near the origin. In other words all
the physics within a region small enough to be considered flat
must be described by the matrix degrees of freedom of the SYM and
not by the spatial variations of the fields. There is nothing
wrong with this except that we have no idea how to translate
ordinary physics into the holographic description. For example we
would have no idea how to determine if a given SYM state were
describing a small ten dimensional black hole, a rock or an
elephant of the same mass.

I would like to suggest that there is a way to do physics which is
complementary to the holographic way but in which bulk phenomena
are much easier to recognize. I would expect that this new way
would be in terms of local bulk fields which would either include
the gravitational field or would allow its construction in some
simple way. What would be unusual about this theory is that it
would be extremely rich in gauge redundancies, so rich in fact
that when the gauge is completely fixed and the  non-redundant
degrees of freedom are counted their number would be proportional
to the area in Planck units. By some particular gauge fixing this
would be made manifest. But after insuring ourselves that the
counting is holographic other gauge choices might be much better
for recognizing ordinary local physics. The kind of theory I have
in mind is some generalization of Chern Simons theory which does
have the property that the real states live on the boundary.
Unfortunately this is just a speculation at the moment.

\newpage

\epsfxsize=150mm  
\epsfbox{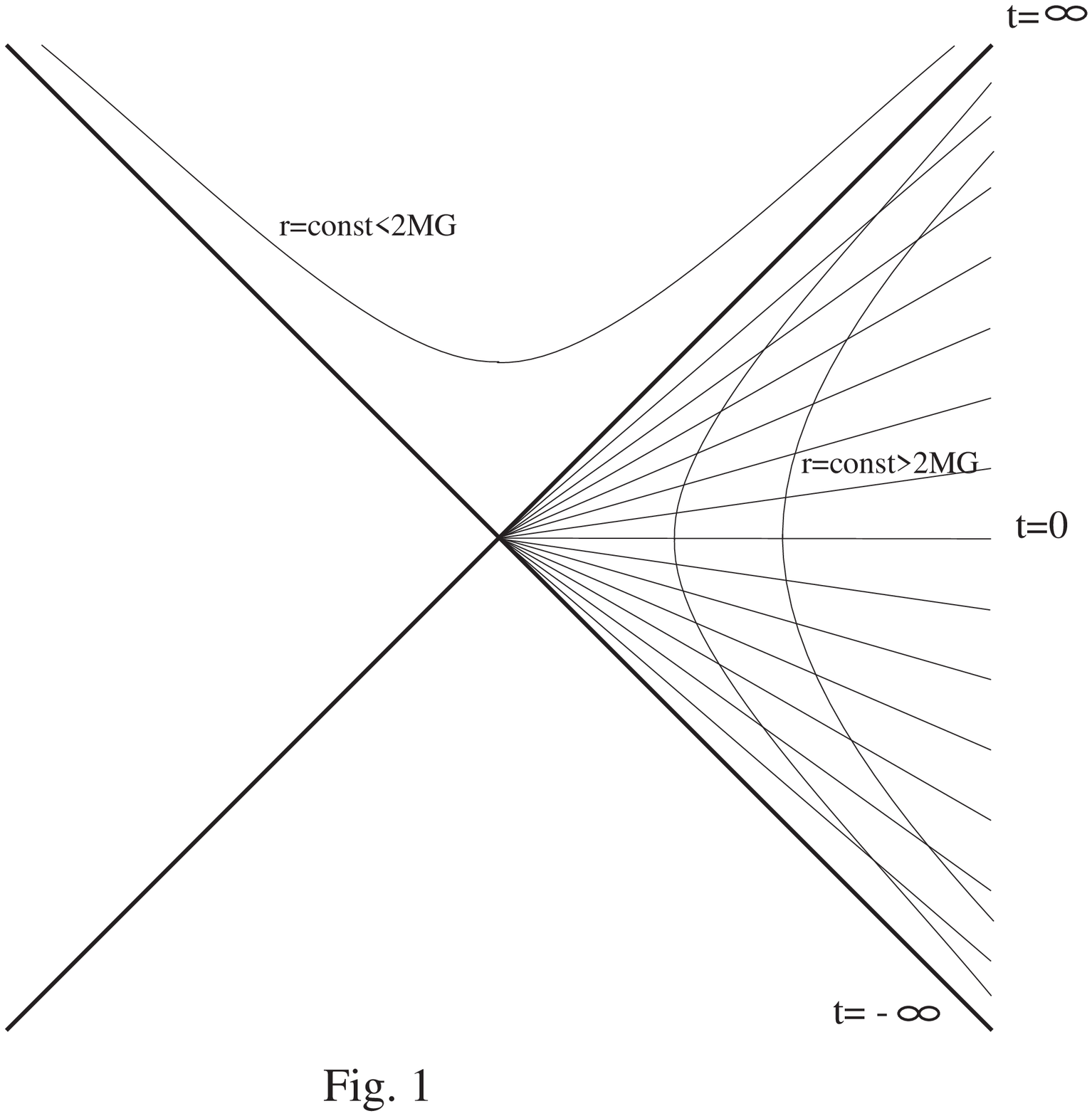} 

\newpage

\epsfxsize=150mm  
\epsfbox{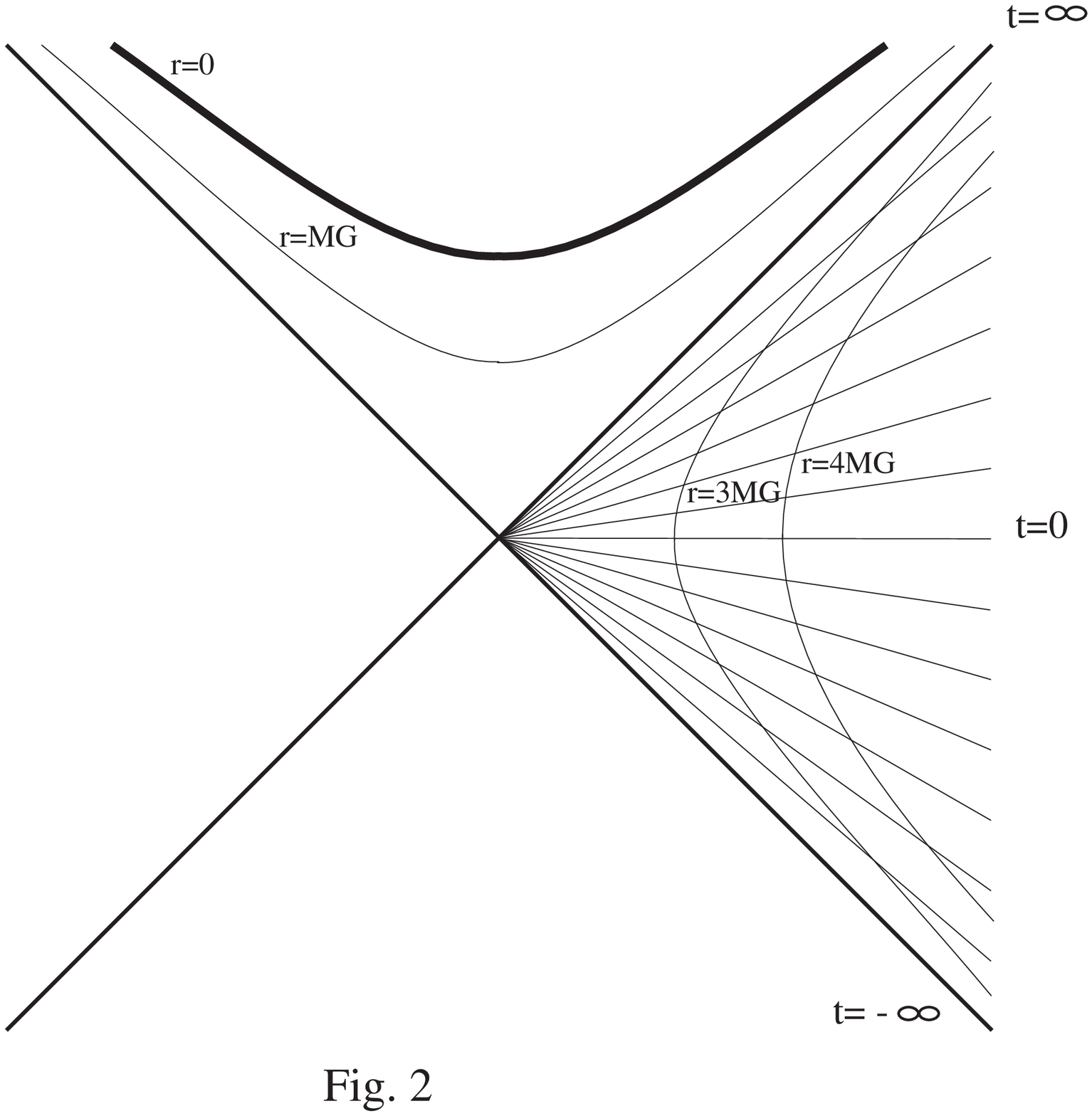} 

\newpage

\epsfxsize=150mm  
\epsfbox{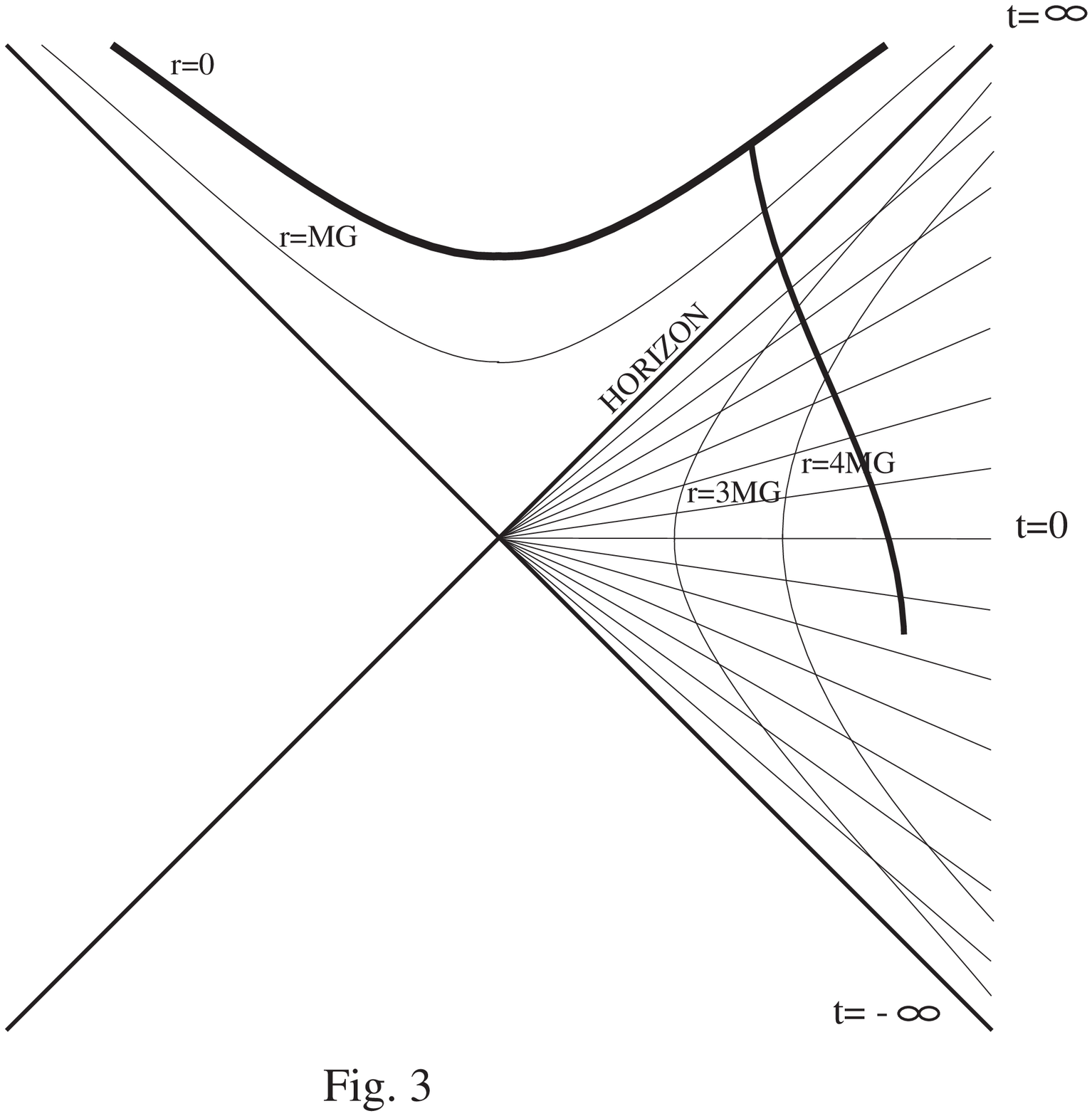} 

\newpage

\epsfxsize=100mm  
\epsfbox{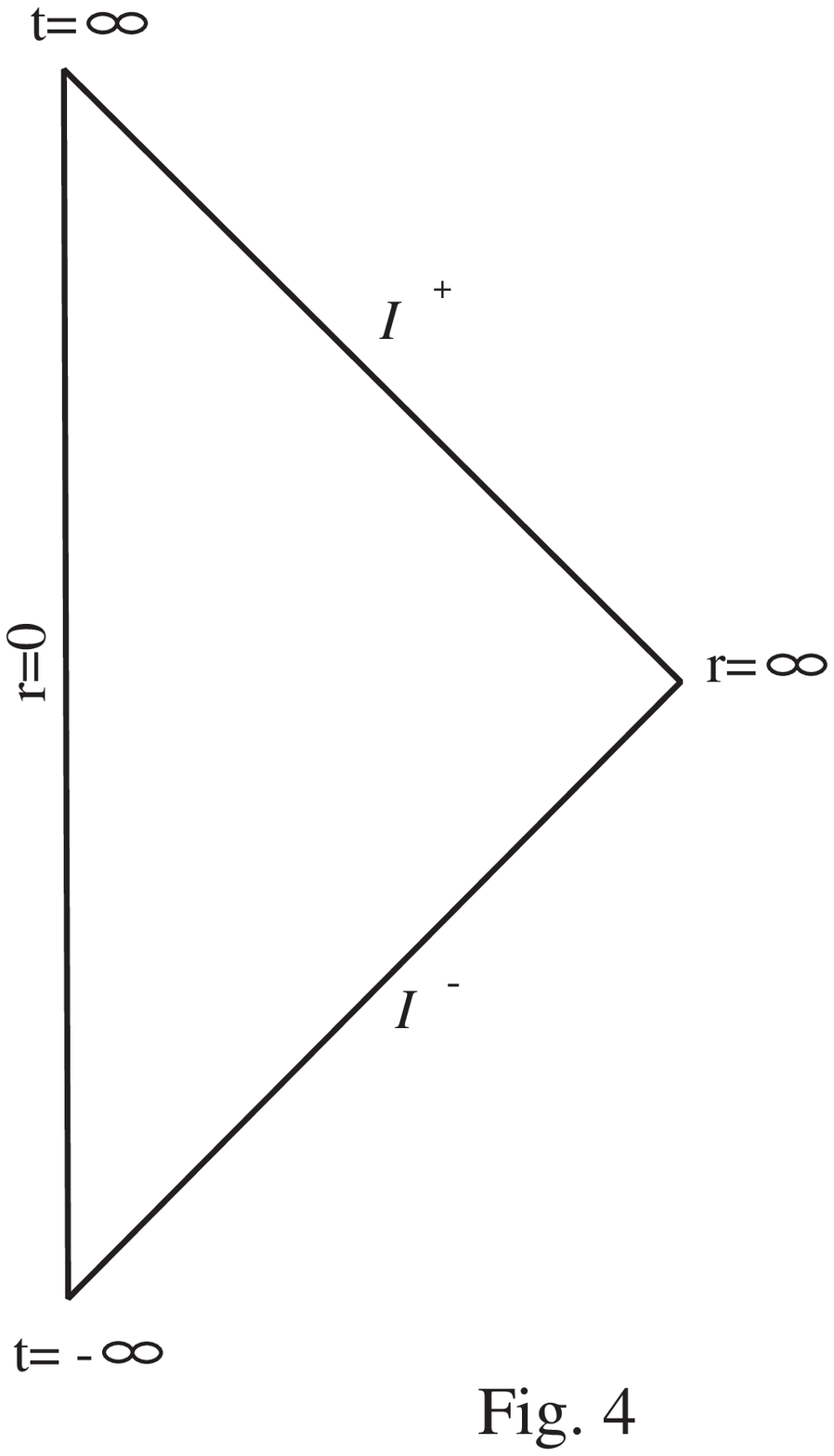} 

\newpage

\epsfxsize=150mm  
\epsfbox{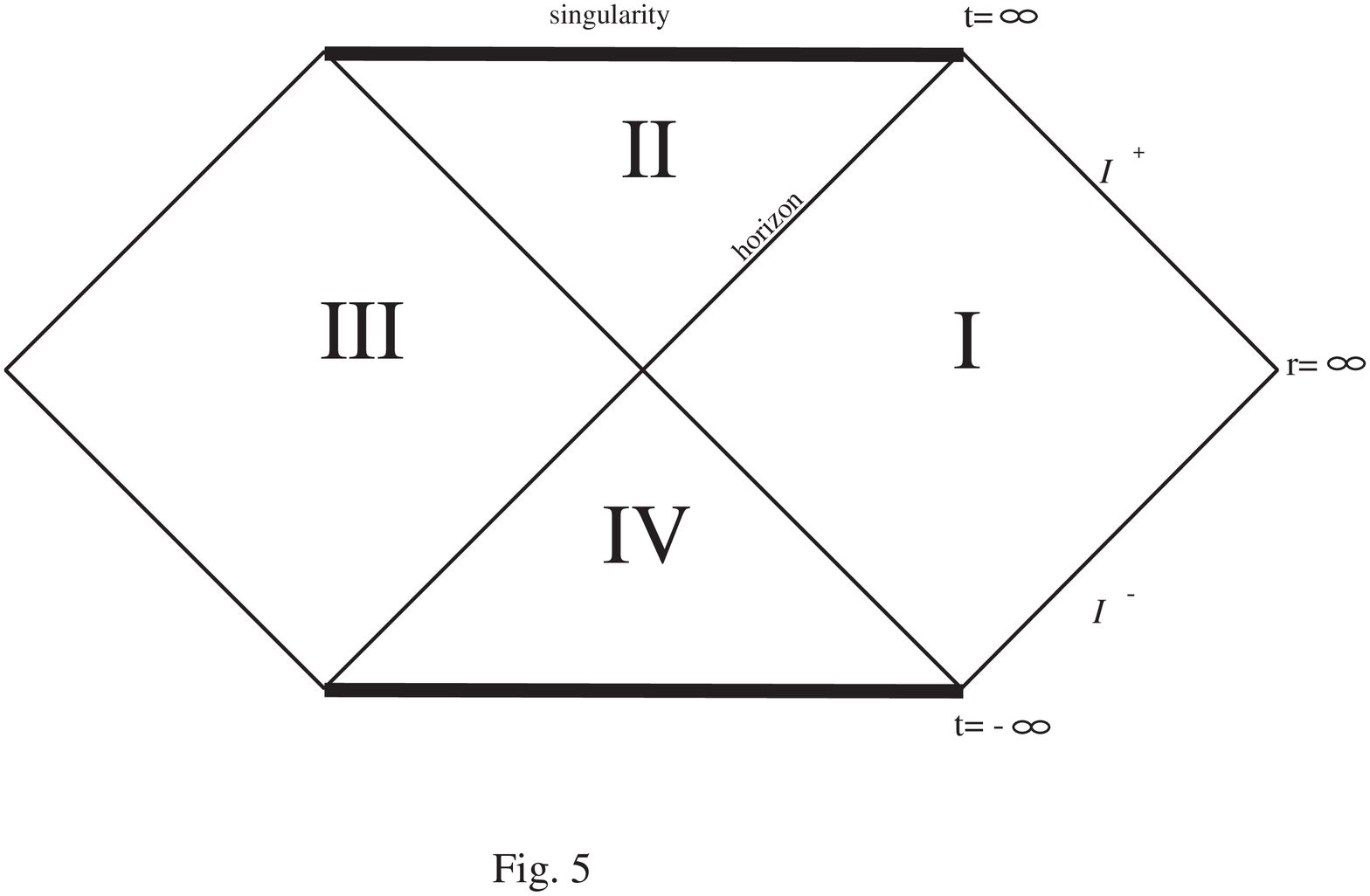} 

\newpage

\epsfxsize=150mm  
\epsfbox{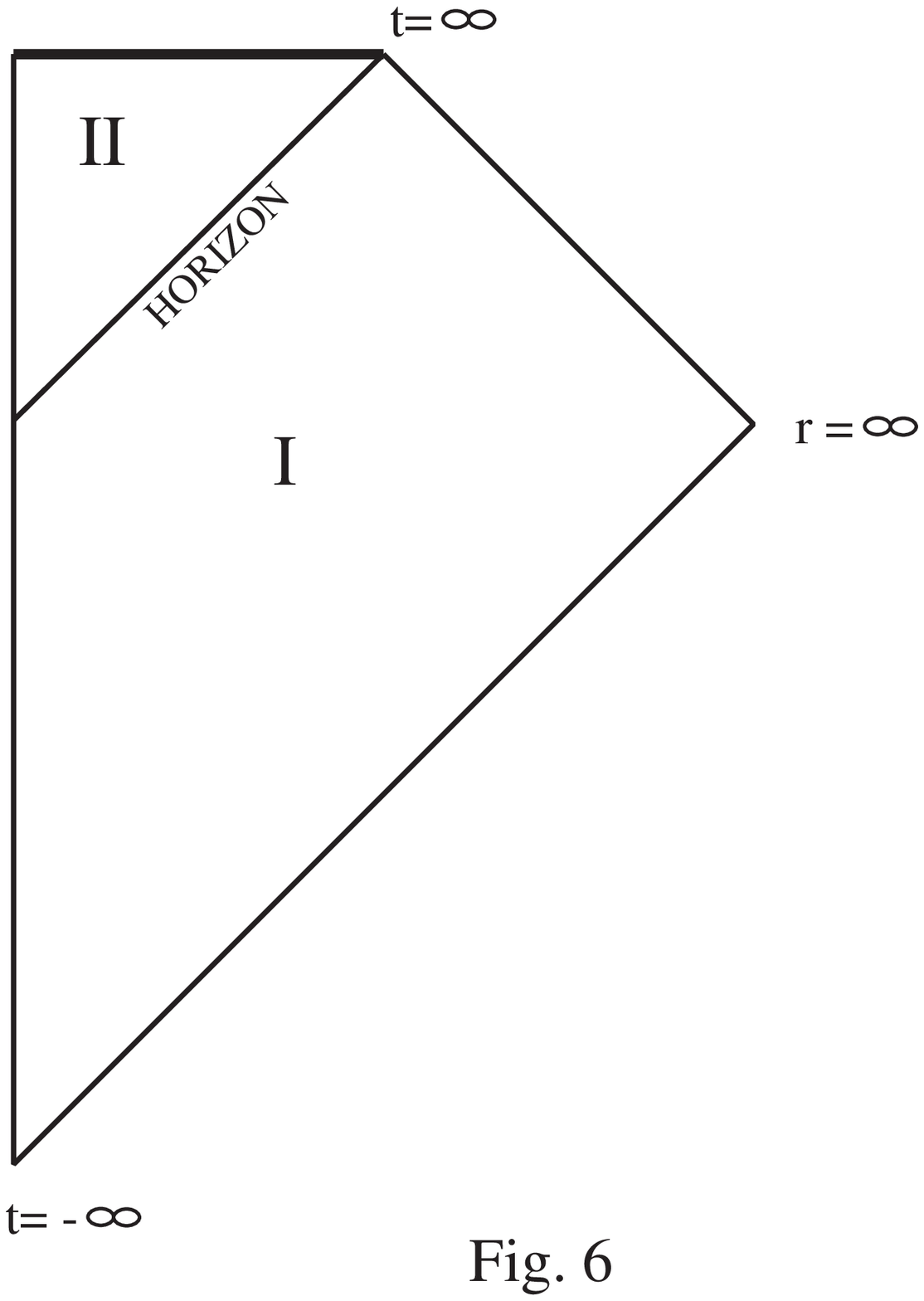} 

\newpage

\epsfxsize=150mm  
\epsfbox{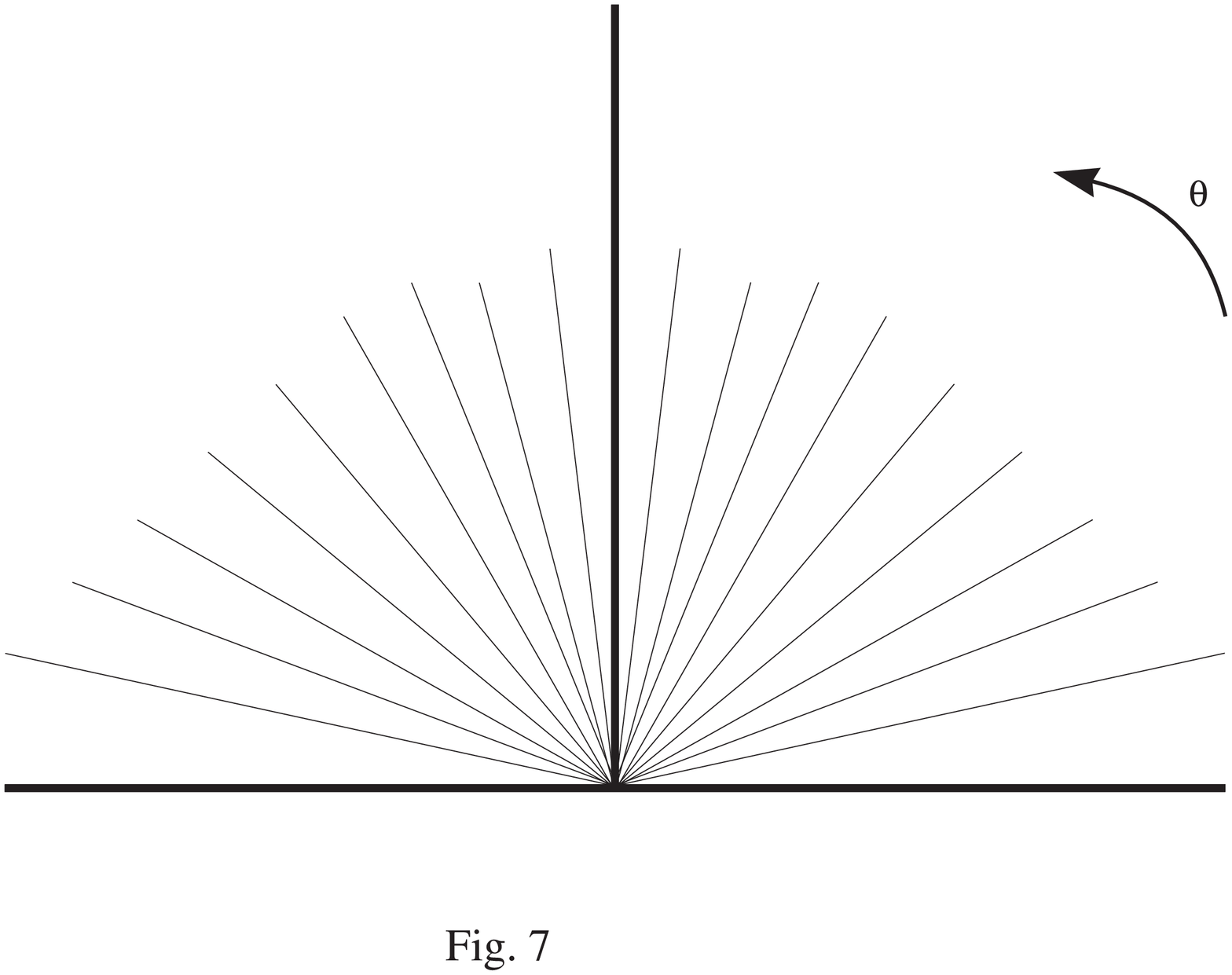} 

\newpage

\epsfxsize=150mm  
\epsfbox{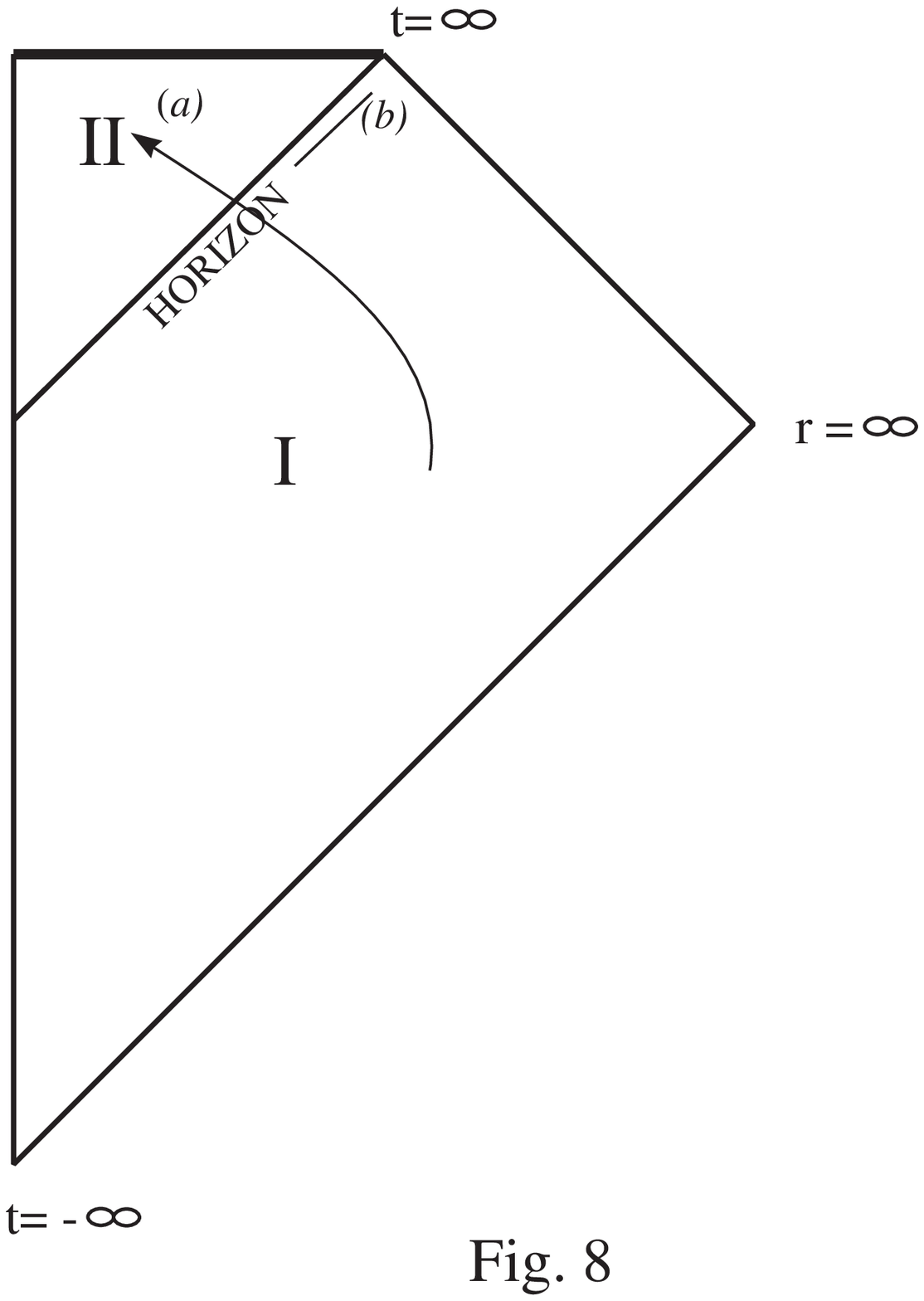} 

\newpage

\epsfxsize=150mm  
\epsfbox{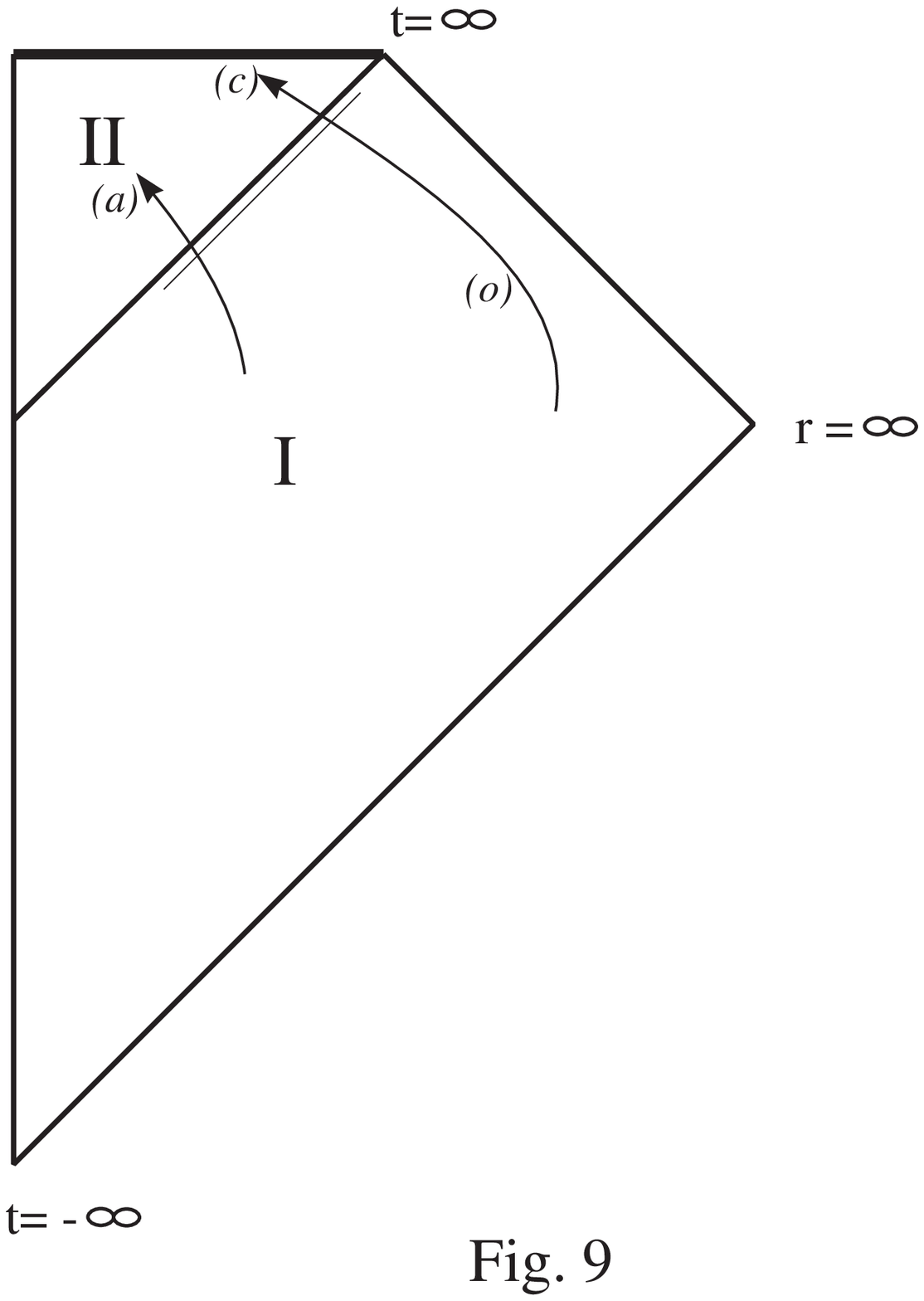} 

\newpage

\epsfxsize=150mm  
\epsfbox{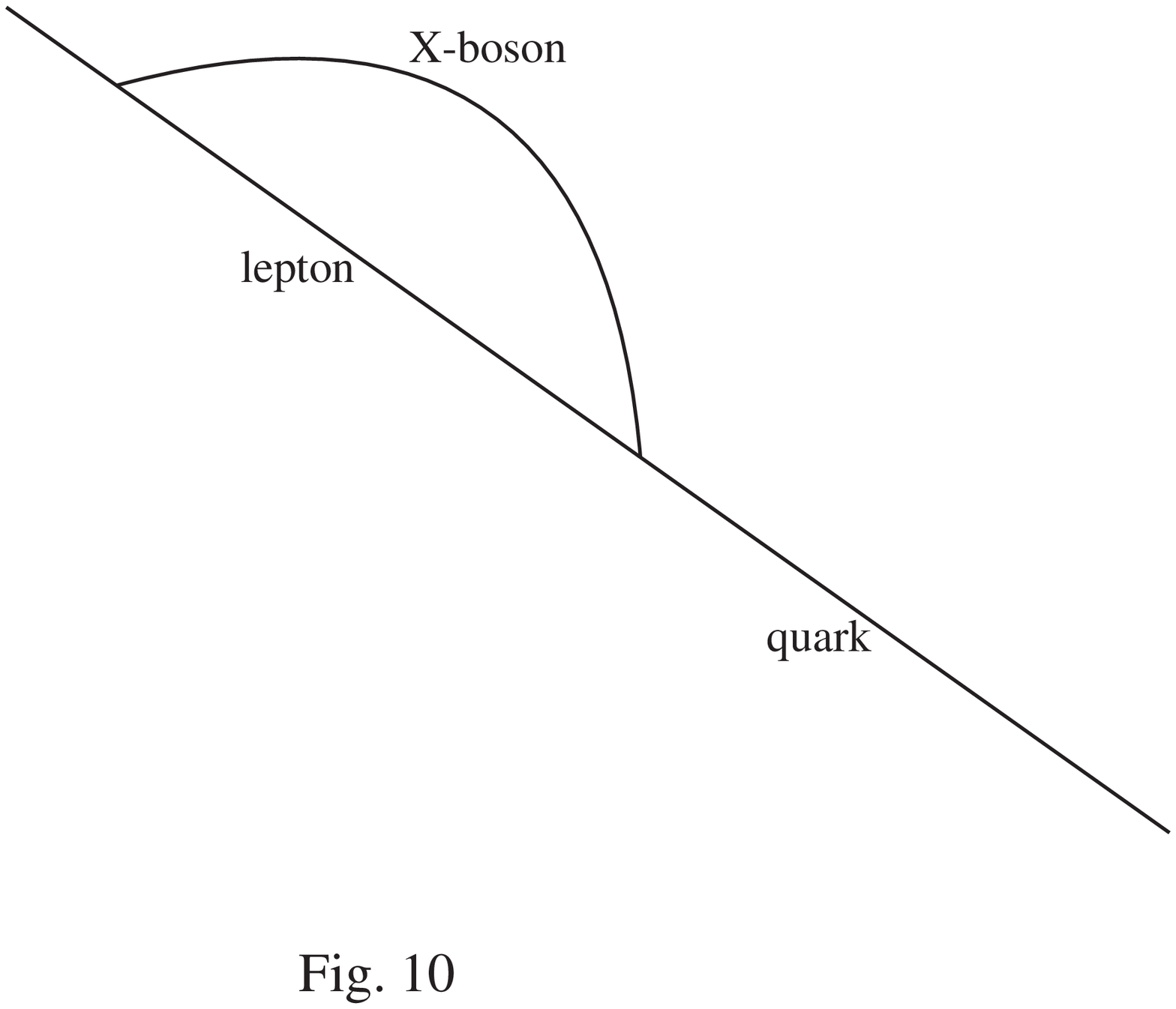} 

\newpage

\epsfxsize=150mm  
\epsfbox{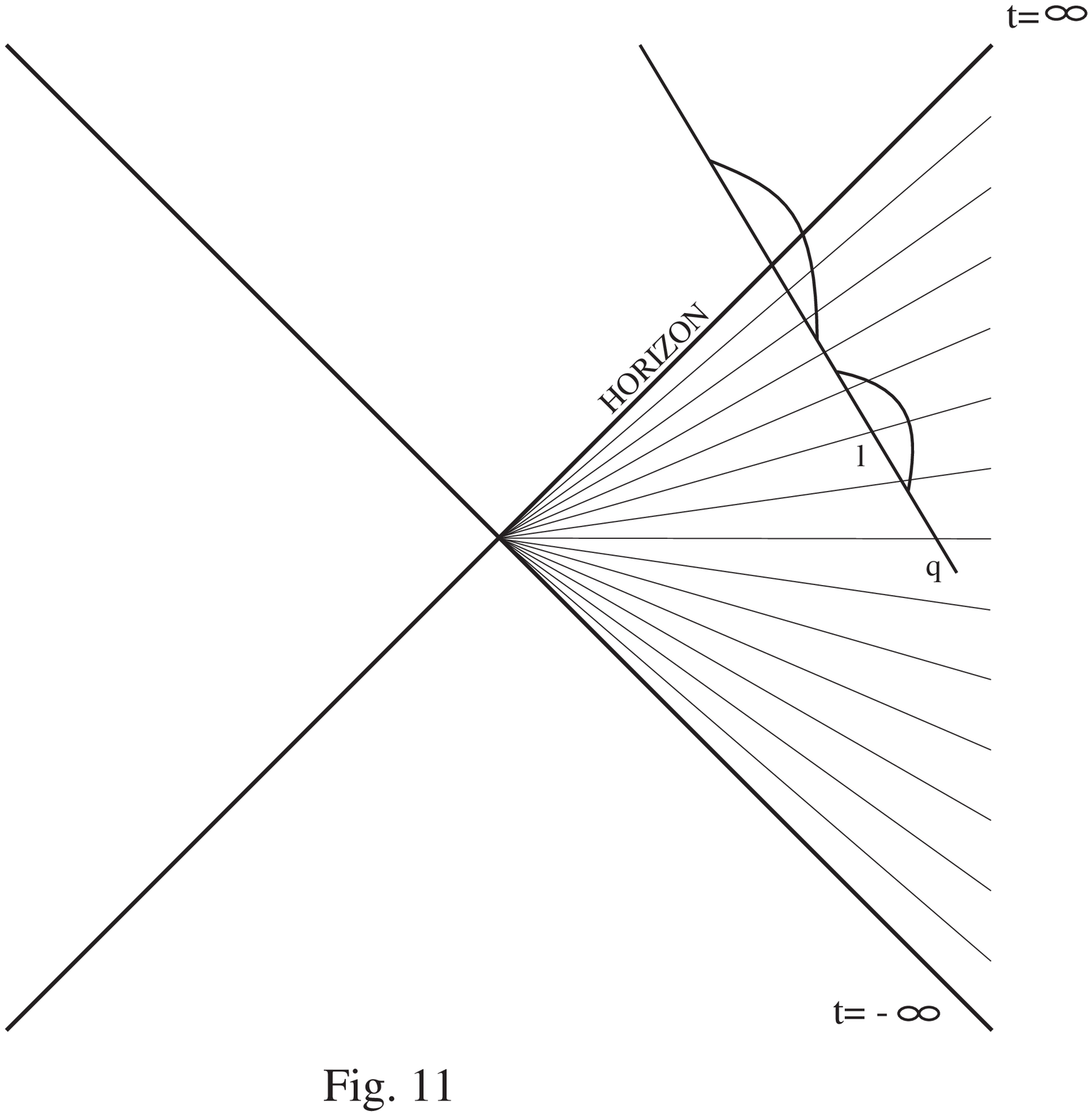} 

\newpage

\epsfxsize=150mm  
\epsfbox{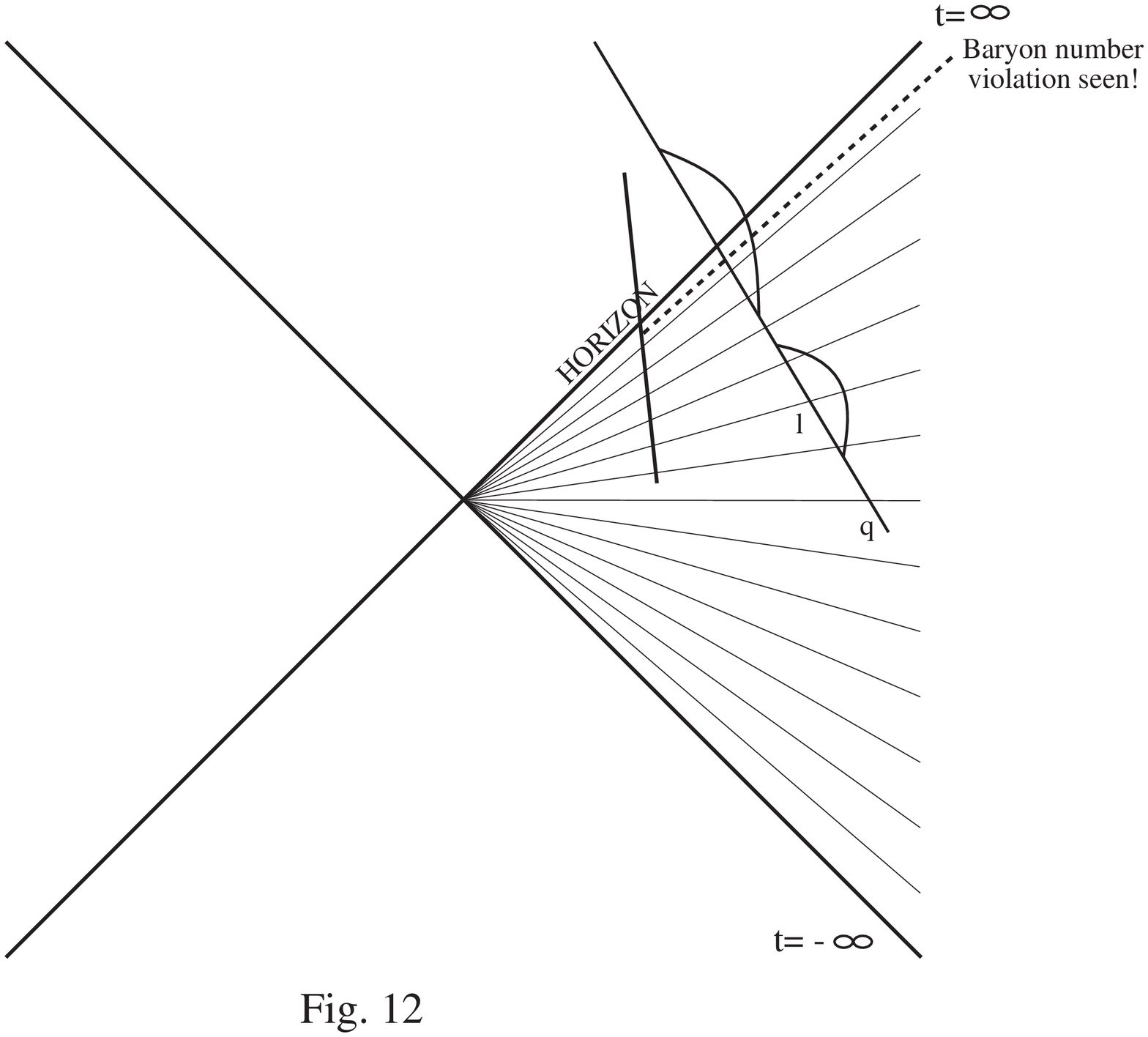} 

\newpage

\epsfxsize=150mm  
\epsfbox{figura13edit2.eps} 

\newpage

\epsfxsize=150mm  
\epsfbox{figura14edit2.eps} 

\newpage

\epsfxsize=150mm  
\epsfbox{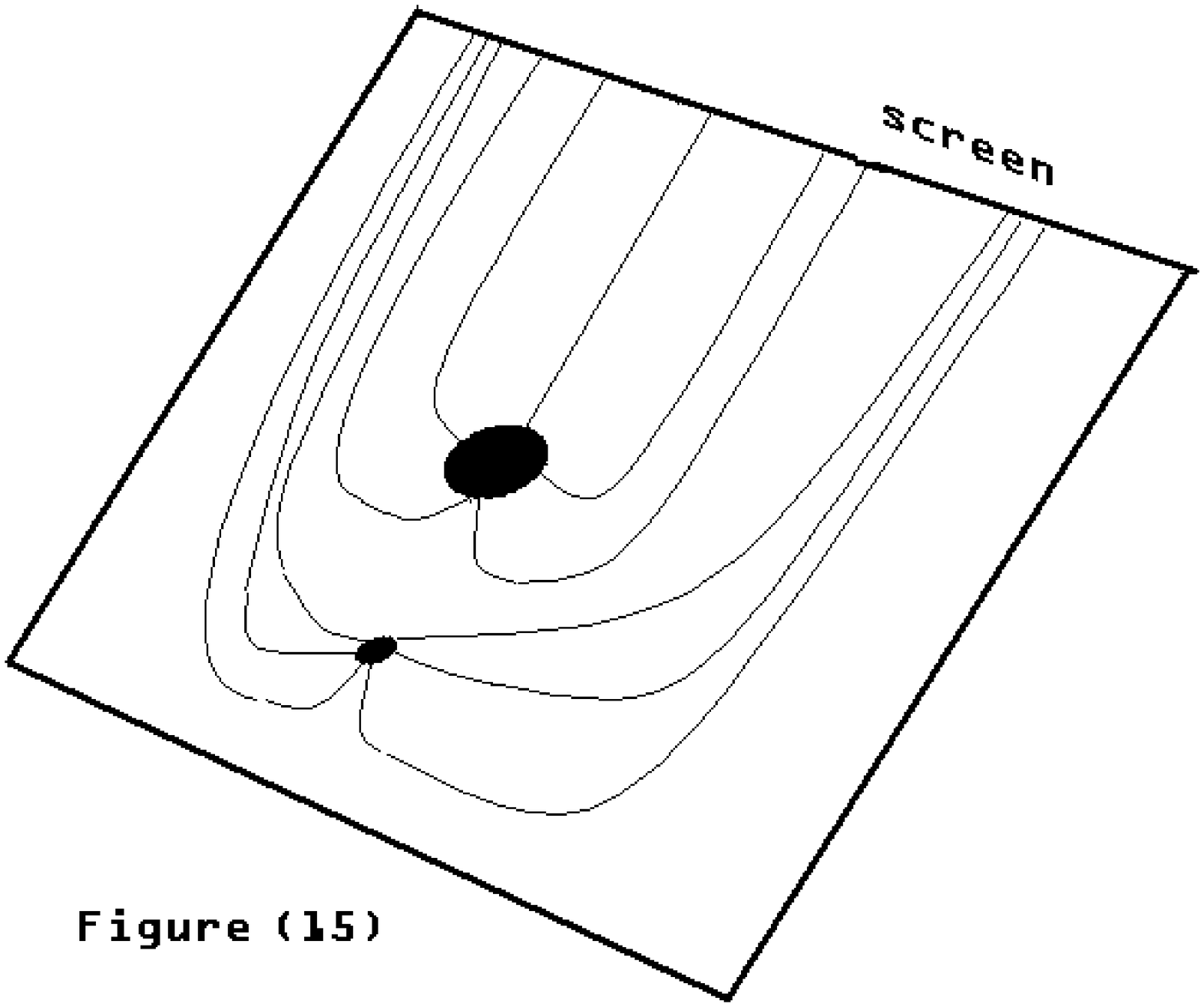} 

\newpage

\epsfxsize=150mm  
\epsfbox{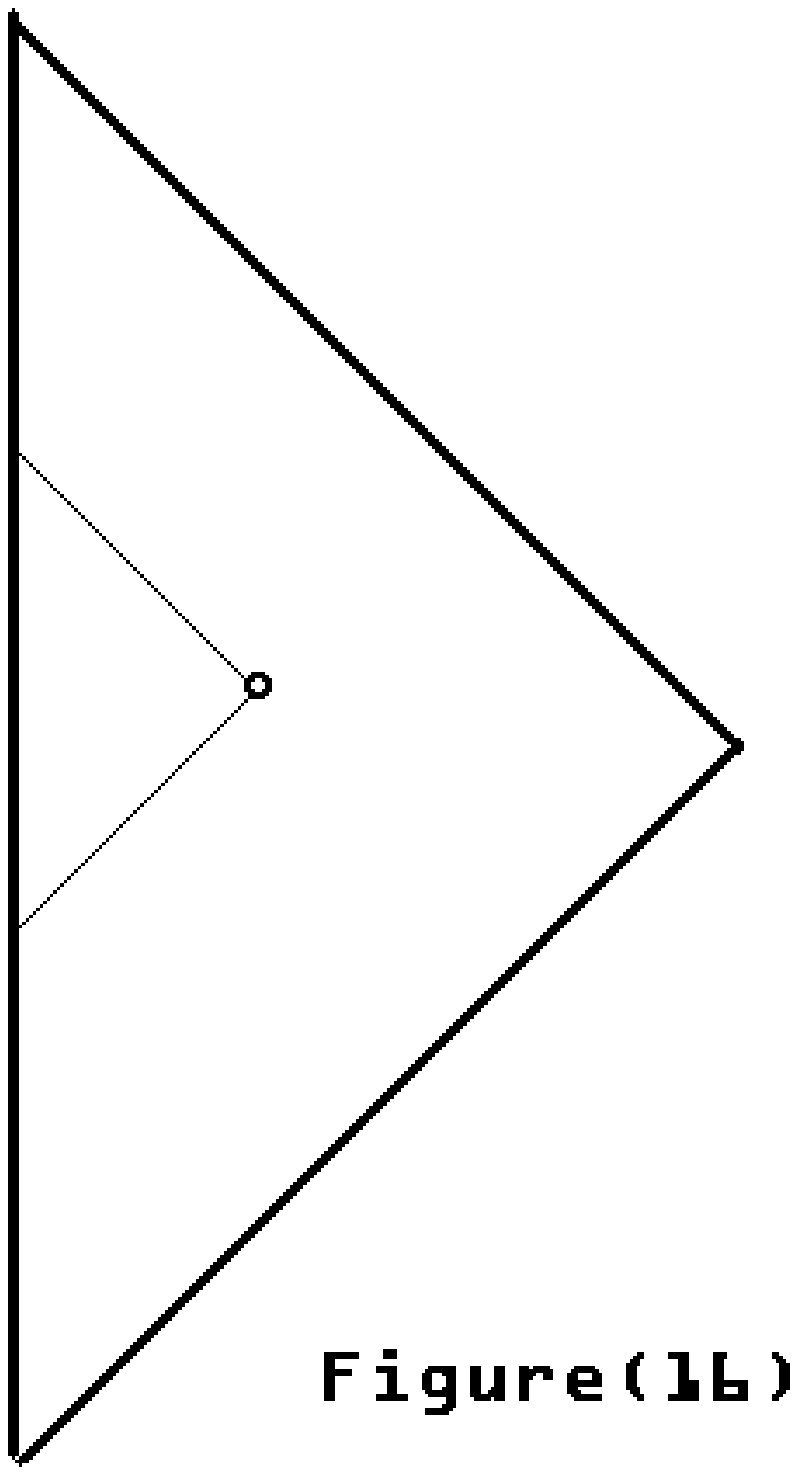} 

\newpage

\epsfxsize=150mm  
\epsfbox{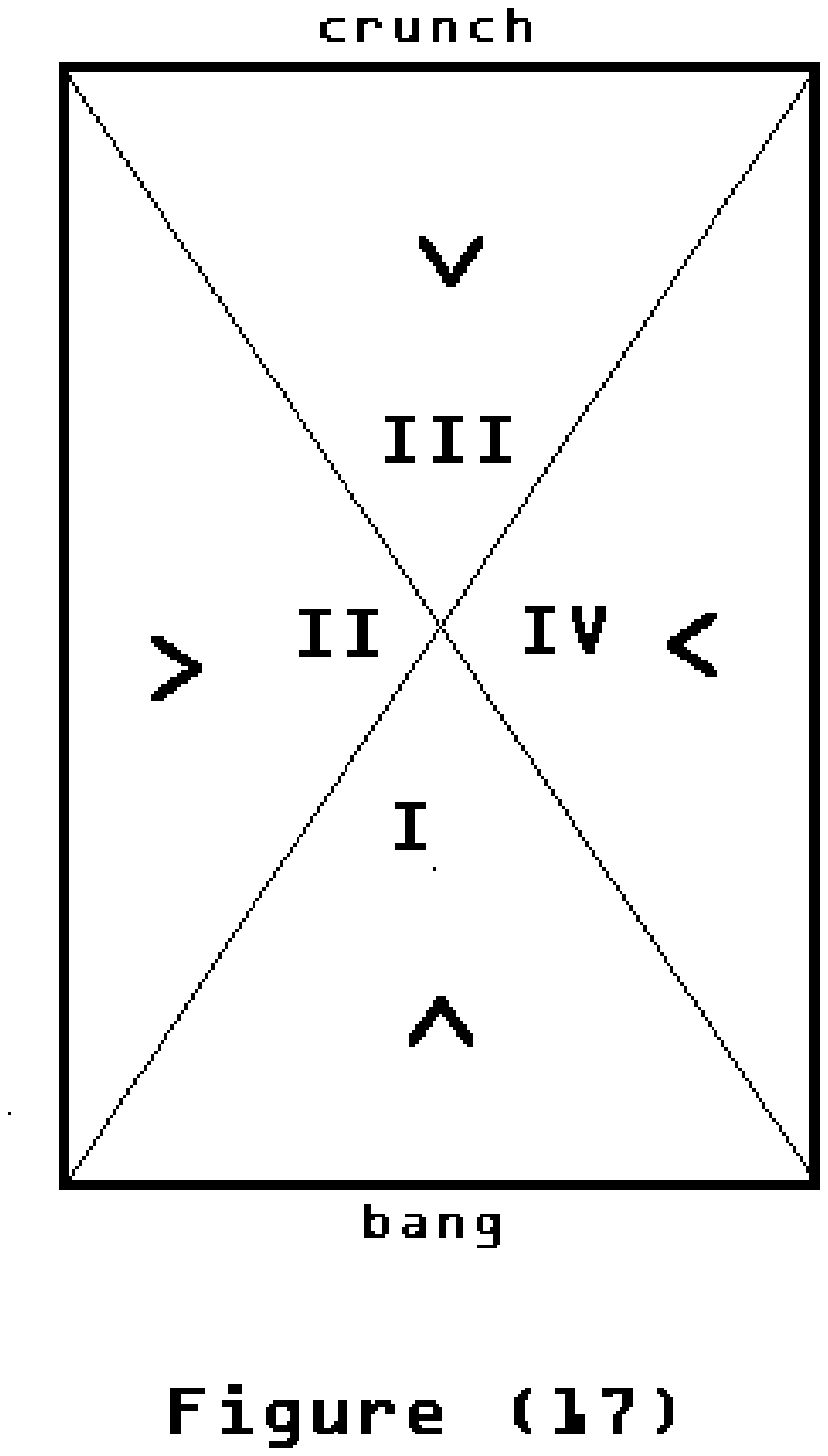} 

\newpage

\epsfxsize=150mm  
\epsfbox{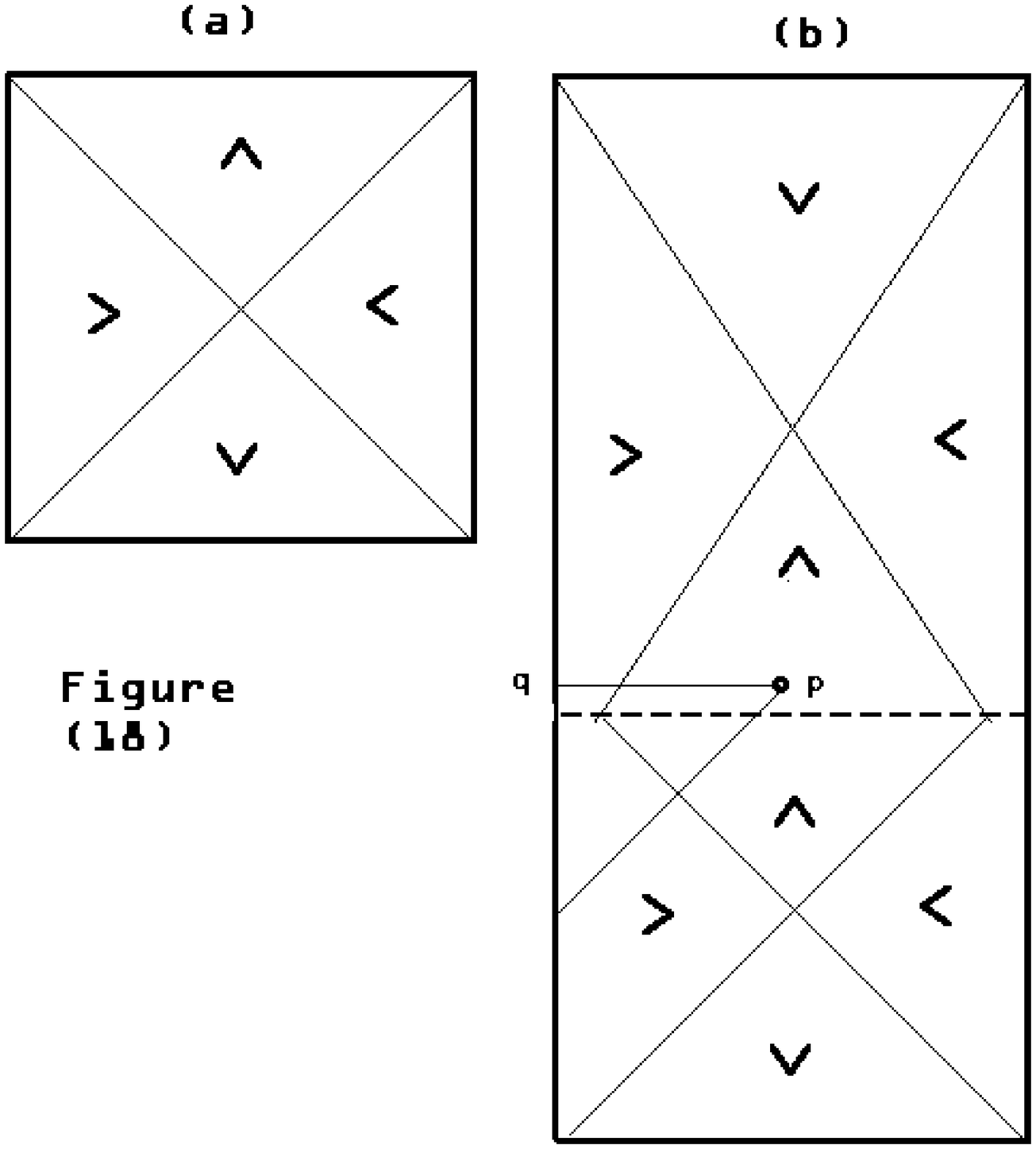} 

\newpage

\epsfxsize=150mm  
\epsfbox{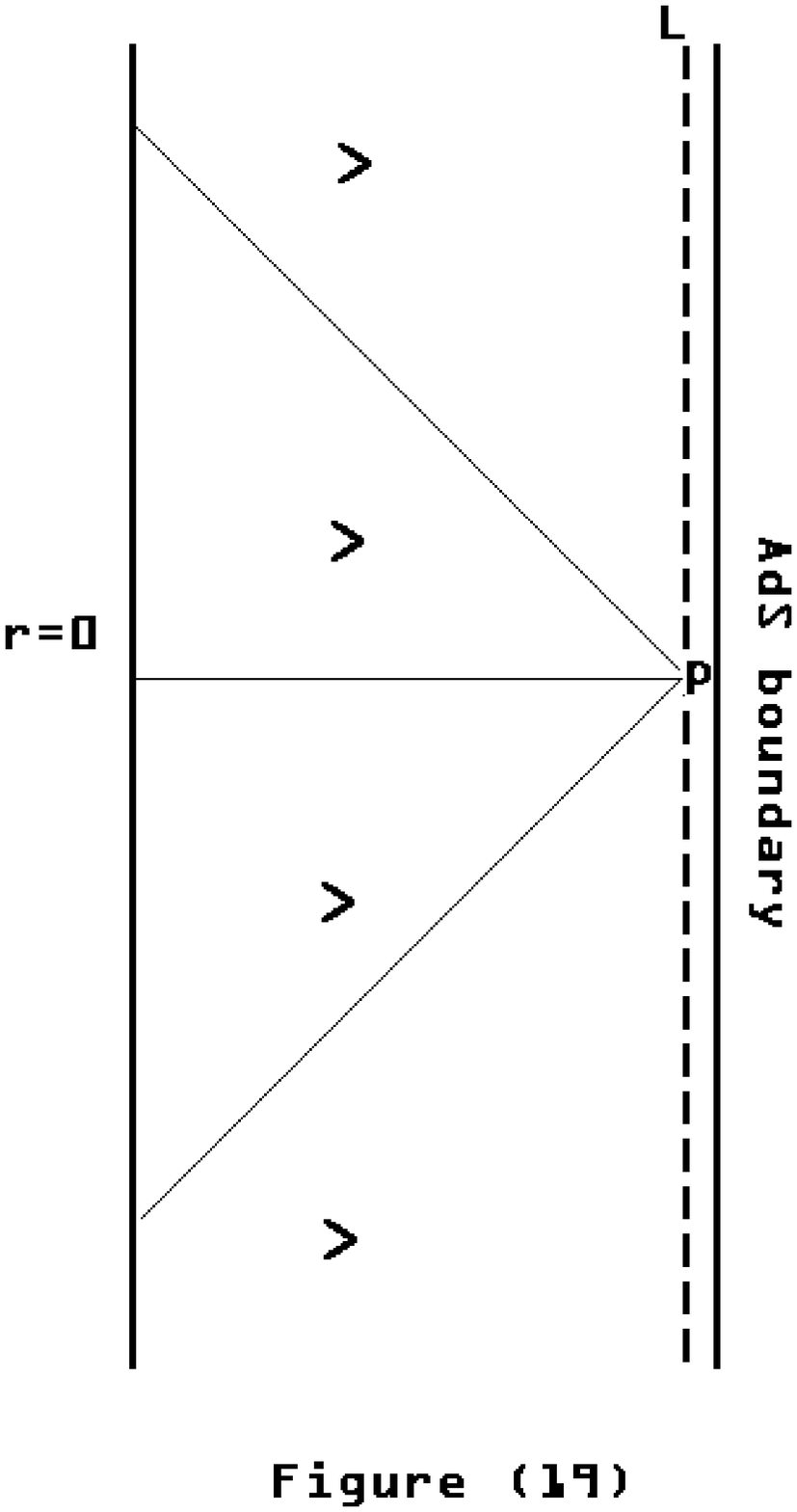} 

\end{document}